\documentstyle[12pt]{article}
\catcode`\@=11
\def\fileversion{1.2a}
\def\filedate{30 Nov 90}
\def\docdate {26 Feb 90}
\wlog{Style option `A4' \fileversion\space<\filedate> (NP and JLB)}
\wlog{English documentation\space\space\space<\docdate> (JLB)}
\ifcase \@ptsize
\or
\or
\fi
\advance\textheight by \topskip
\ifcase \@ptsize
  \if@twoside
  \else
  \fi
\or
  \if@twoside
  \else
  \fi
\or
  \if@twoside
  \else
  \fi
\fi
\def\WideMargins{%
  \newdimen\ExtraWidth
  \ifcase \@ptsize
    \ExtraWidth = 0.5in
    \@widemargins
  \or
    \ExtraWidth = 0.5in
    \@widemargins
  \or
    \ExtraWidth = 0.7in
    \@widemargins
  \fi\let\WideMargins\relax\let\@widemargins\relax}
{\def\do{\noexpand\do\noexpand}
 \xdef\@preamblecmds{\@preamblecmds \do\WideMargins}
}
\def\@widemargins{%
    \global\advance\textwidth by -\ExtraWidth
    \global\advance\marginparwidth by \ExtraWidth
    \if@twoside
      \tw@sidedwidemargins
    \else
      \@nesidedwidemargins
    \fi}
\def\tw@sidedwidemargins{%
    \if@reversemargin
      \@tempdima=\evensidemargin
      \advance\@tempdima by -\oddsidemargin
      \advance\oddsidemargin by \ExtraWidth
      \advance\oddsidemargin by \@tempdima
      \advance\evensidemargin by -\@tempdima
    \else
      \advance\evensidemargin by \ExtraWidth
    \fi}
\def\@nesidedwidemargins{%
    \if@reversemargin
      \advance\oddsidemargin by \ExtraWidth
      \advance\evensidemargin by \ExtraWidth
    \fi}
\ifcase \@ptsize
\or 
\or 
\fi
\catcode`\@=12
\def\theequation{\arabic{section}.\arabic{equation}}

\begin{document}

\begin{titlepage}
\title{Semiclassical equations for weakly inhomogeneous cosmologies}
\author{\\ \\ \\  Antonio Campos and Enric Verdaguer \\
       {\normalsize\it Grup de F\'{\i}sica Te\`orica and I.F.A.E.,} \\
       {\normalsize\it Universitat Aut\`onoma de Barcelona,} \\
       {\normalsize\it 08193 Bellaterra (Barcelona), Spain} }
\date{}
\maketitle

\vskip-12cm\hskip13cm UAB-FT 316
\vskip0.2cm\hskip13.25cm JUNE 1993
\vskip 12cm

\begin{abstract}
The in-in effective action formalism is used to derive the
semiclassical correction to Einstein's equations due to a massless
scalar quantum field conformally coupled to small gravitational
perturbations in spatially flat cosmological models. The vacuum expectation
value of the stress tensor of the quantum field is directly derived from the
renormalized in-in effective action. The usual in-out effective action
is also discussed and it is used to compute the probability  of particle
creation. As one application, the stress tensor of a scalar field around
a static cosmic string is derived and the backreaction effect on the
gravitational field of the string is discussed.
\end{abstract}

\end{titlepage}

\section{Introduction}
\setcounter{equation}{0}

Our picture of the evolution of the early universe relies in the so
called semiclassical theory of gravity which describes the
interaction of quantum fields with the classical gravitational
field.  Order of magnitude arguments and Heisenberg's uncertainty principle
tell us that there must be a period in the universe evolution, well after the
Planck time, when the quantization of the gravitational field may be ignored
but still the scale of its time variations is short enough to create
elementary particles, so that matter quantization cannot be ignored.
Since we lack a theory of quantum gravity  it
is still not known to what extent and in what sense this theory may be
considered as a true semiclassical limit of quantum gravity
interacting with matter fields. Plausibility arguments have been advanced by
Hartle and Horowitz \cite{HHo} who show that the quantum corrections to the
classical action of gravity interacting with $N$ identical non
self-interacting
matter fields reduce in the leading-order $1/N$ approximation to such
semiclassical theory.

The semiclassical approach provides the framework for some
realistic scenarios which
may explain some of the features of the present universe. One of these
scenarios is inflation \cite{Inf} which may explain the
homogeneity and flatness problems of the standard big bang cosmology. In the
inflationary model the quantum fluctuations of the inflaton field may be the
source of the small gravitational inhomogeneities which seed galaxies or
gravitational waves. This may
explain the universe large scale structure \cite{Strfor} and the presence
of a hypothetical background of gravitational radiation \cite{Gri75-93}.
Another scenario is the possible formation of topological defects
\cite{Topdef} as the universe undergoes some phase transitions.
Topological defects, in particular cosmic strings, may seed
structure \cite{costr} and may be an alternative to inflation
for the generation of structure in the universe.

In both scenarios the picture of the gravitational field that emerges is that
of a conformally flat Friedmann-Robertson-Walker (FRW) background in
which small gravitational perturbations are present. Large anisotropies and
inhomogeneities might be present only if the universe had emerged highly
inhomogeneous from the Planck era into the classical regime
\cite{anis}. But since the quantum consequences of a highly
inhomogeneous model are difficult to estimate one assumes that by studying
small perturbations on a FRW background, a qualitative picture
of the evolution of the more extreme case may result.

Here we are interested in the quantum effects produced by the presence of
small perturbations in conformally flat backgrounds. Quantum
effects due to small anisotropies were first considered by Zeldovich
\cite{Zel70}, Zeldovich and Starobinsky \cite{ZS72}, Hu and Parker
\cite{HP}, Hartle and Hu \cite{HH}, and Birrell and Davies \cite{BD},
who computed the creation of conformally coupled particles interacting with
the
anisotropies. Conformally coupled particles are not created in conformally
flat backgrounds (FRW) \cite{Par68} but the anisotropies break the
conformal symmetry. Different techniques were used for such computations,
these techniques go from a perturbative evaluation
of the Bogoliubov transformations
relating two vacua of the quantum field, to the evaluation of the in-out
effective action of this field in the given gravitational background.
These results were extended to the presence of arbitrary perturbations,
including inhomogeneities, \cite{inh} by a technique
based in the perturbative evaluation  of the scattering matrix which had been
used  in flat backgrounds by Sexl and Urbantke \cite{SU} and Zeldovich and
Starobinsky \cite{ZS77}.

Quantum effects on the geometry,
the so called backreaction effect, are more difficult to evaluate because this
requires, on the one hand, the computation of the renormalized stress tensor
of the quantum field in order to modify the classical Einstein equations, and
on the other hand, it requires the solution to these
semiclassical equations. It was argued by
Zeldovich \cite{Zel70} that the backreaction would tend to dissipate the
inhomogeneities as in a sort of gravitational Lenz's law effect.
This is a mechanism to homogeneize the universe, but it is usually
not advocated because in the standard scenario one assumes
cosmological models which cannot explain the present large scale
homogeneity by any causal mechanism after the Planck era;
the inflationary scenario, on the other hand,
seems to solve the homogeneity problem quite naturally.
It is nevertheless a mechanism for entropy production. Early work on
the backreaction effect on the geometry due to anisotropies was done by Lukash
and Starobinsky \cite{LS74} and Lukash et al. \cite{Lu76}, who assumed very
special conditions near the Planck time, and by Hu and Parker \cite{HP} who
considered a Bianchi type I anisotropic model, evaluated the stress tensor in
the low frequency approximation and computed the resulting modified Einstein's
equations numerically. The results of such work indicate that the dynamical
mechanism of particle production achieves a rapid damping of the anisotropy if
the calculations are extrapolated to the Planck era.

The computation of the quantum stress tensor is generally difficult in
practice. However, for small perturbations on a conformally flat background
one may use perturbative methods to get explicit expressions. One of the most
powerful and efficient methods and, one that is very well adapted to a
perturbative scheme, is based on the one loop order computation of the so
called in-in effective action for quantum fields interating with the
gravitational perturbations. This technique is an effective action technique
adapted to compute expectation values of quantum operators. It was first
proposed by Schwinger \cite{Sch} and Keldysh \cite{Kel} and developed by Chou
et al. \cite{Cho}. Jordan \cite{Jor86} and Calzetta and Hu \cite{CH}
developed the technique on a curved background, and it was then applied to
derive the stress tensor of a quantum scalar field coupled to small
anisotropies on a cosmological background \cite{CH}.

The use of effective action methods in the backreaction context was first
considered by Hartle \cite{Har}, and Fischetti et al. \cite{FHH}, and
Hartle and Hu \cite{HH} studied the effect of anisotropies. But in their
formalism the basic element is
the usual effective action which is related to the generating functional of
the
in-out vacuum persistence amplitude. This in-out formalism leads to matrix
elements rather than expectation values for the quatum operators. Thus,
One does not get directly from the in-out effective action the vacuum
expectation value of the stress tensor of the quantum field, and
one still needs to compute the Bogoliubov transformation between the in and
out vacua. This method is, however, very useful for the computation of the
particles created, since the probability amplitude for particle creation is
directly related to the vacuum persistence amplitude.

In this paper we compute the in-in effective action to the one loop order
for a massless scalar field conformally coupled to small gravitational
perturbations on a spatially flat FRW background. The in-in effective
action is used to derive the quantum stress tensor and the
corresponding semiclassical Einstein's equations. Our
results generalize the Calzetta and Hu \cite{CH} results to the case of
arbitrary small perturbations including inhomogeneities,
and the stress tensor we derive
coincides with that obtained by Horowitz and Wald \cite{HW}, who used an
axiomatic approach to derive it, and by Starobinsky \cite{Sta81} who used a
modified Pauli-Villars regularization method \cite{ZS72}.
One should stress, however, that the stress
tensor computed does not include the energy of the particles created which
is a
second order correction to the computed terms; it includes only first order
vacuum polarization effects. One might wonder that although the energy of the
particles created is small it might have a long term cumulative effect.

We should mention that the axiomatic approach to derive the stress tensor has
been quite succesful in several situations.
Thus Horowitz \cite{Hor80} obtained
the stress tensor due to a scalar field minimally coupled to arbitrary
gravitational linear perturbations on a flat spacetime background applying the
axiomatic arguments outlined by Wald \cite{Wal77}.
This tensor was rederived by
Jordan \cite{Jor87} using the in-in effective action method; note that the
case of conformally coupled fields may also be obtained from our
cosmological model when the conformal factor is taken constant. Another
approach to the quantum stress tensor based on an iteratively evaluated mode
decomposition was developed by Davies and Unruh \cite{DU}.

In this paper we shall not consider the solutions to the semiclassical
equations, except in  a simple example involving a
cosmic string. The correct approach to this problem is still controversial. In
fact, the semiclassical equations are known to admit runaway solutions as a
consequence of the fact that they are dynamical equations with higher order
derivatives. Horowitz \cite{Hor80} and Jordan \cite{Jor87}
found from these solutions that flat
space is unstable against quantum effects. Whether these solutions are
physical and thus signal a true instability, or unphysical and thus spurious,
has been the subject of some
discussion in recent years. Simon \cite{Sim91} has
argued that the semiclassical correction to Einstein's equations must be seen
as analytic perturbations, in terms of the Planck constant $\hbar$, to the
classical Einstein's equations and that,
as such, only solutions which are also
analytic in $\hbar$ are physical. A consistent perturbative approach to find
reduced equations, i.e. dynamical equations which are second order at each
order of perturbation is known \cite{JLM}. When this is applied
it is found that flat
space is perturbatively stable to first order in $\hbar$ \cite{Sim91}. Reduced
semiclassical equations have been obtained also in some cosmologies \cite{PS}.
Suen \cite{Sue}, on the other hand, has argued, on the basis of how the
stress tensor is renormalized, that this tensor cannot be considered
the first term of an expansion and therefore the previous reduction methods
should not be applied.

In order to make this paper reasonably self-contained the in-in effective
action formulation is summarized in section 2 with a view to practical
applications. In section 3 the in-in effective action
to the one loop order is derived
for a scalar field conformally coupled to a nearly conformally flat metric.
Since along this computation one derives also all the terms needed for the
in-out effective action, this action is also discussed,
and it is used to derive the probability for pair creation;
the results agree with those obtained by other methods. In section 4
the stress tensor for the quantum field is derived from the in-in effective
action and the semiclassical correction to
Einstein's equations is written down. The stress tensor is seen to agree with
that obtained by Horowitz and Wald \cite{HW} and Starobinsky \cite{Sta81}.
As an exercise the semiclassical equations in two dimensional spacetime are
also derived using the same formalism. In
section 5 we apply the previous formula to compute the stress tensor of a
quantum field around a static cosmic string and we discuss the backreaction
effect on the gravitational field of the string. Note that
since the gravitational field of a
cosmic string can be considered a small perturbation on a flat background
the above perturbative technique (in the sense of metric perturbations)
can be applied. The results are in agreement
with those found by other non perturbative methods \cite{His87}\cite{LHK} but
this perturbative method opens the possibility of computing the
quantum stress tensor even in time dependent situations.
Work along these lines is in progress.

\section{In-in functional formalism}
\setcounter{equation}{0}

In this section we summarize the in-in functional formalism for the
evaluation of the in-in effective action with a view to the
applications of this paper. We follow, essentially, the presentations
by Jordan \cite{Jor86}, Calzetta and Hu \cite{CH}, and Paz \cite{Paz}.

Quantum corrections to a classical field theory can be studied with the
help of the effective action. For simplicity, we consider the
quantization of a scalar field $\phi (x)$. The usual in-out
effective action is based in the generating functional
$W[J]$ which is related
to the vacuum persistence amplitude in the presence of some classical
source $J(x)$ by
\begin{equation}
     e^{iW[J]} \equiv \langle 0,out|in,0\rangle_J .
\end{equation}
This functional carries all the quantum
information of the connected graphs of the theory.

When one couples an external field $J(x)$ it is convenient to use the
interaction picture in which the
states $|\psi\rangle$ evolve in time according
to the Schr\"odinger equation
$H_I|\psi\rangle=i\partial_t |\psi\rangle$, where
$H_I$ is the interaction Hamiltonian
operator $H_I=\int d^{n-1}x J(x)\phi (x)$,
$\phi (x)$ is now the field operator in the Heisenberg representation
and $n$ the number of spacetime dimensions.
The solution of this equation may be
formally written as,
\begin{equation}
     |\psi\rangle_{t_2}= T^{(t)}\exp\left( i\int_{t_1}^{t_2}
             dt H_I\right) |\psi\rangle_{t_1},
\end{equation}
where $T^{(t)}$ is the usual time ordering operator, and (2.1) can be written
as
\begin{equation}
     e^{iW[J]}= \langle 0,out| T^{(t)}\exp\left(i\int_{-\infty}^{\infty}
                dt H_I\right) |in,0\rangle .
\end{equation}
It is easy to see from the classical field equations for $\phi(x)$ in the
presence of $J(x)$ that $\exp iW[J]$ satisfies the integro differential
Schwinger-Dyson equation, and that one may give a path integral
represenation for its solution as,
\begin{equation}
     e^{iW[J]}= \int {\cal D} [\phi] e^{i(S[\phi] + J\phi)} ,
\end{equation}
where $S[\phi]$ is the classical action of the field theory and the
common shorthand notation $J\phi$ for the integral
$\int d^nxJ(x)\phi(x)$ has been used. The functional integral is taken with
the following boundary conditions: $\phi\rightarrow e^{\mp i\omega t}$,
where $\omega > 0$, when the time $t\rightarrow\pm\infty$,
$i.e.$ the scalar field has only negative and positive
frequency modes in the in and out regions respectively; the interaction is
assumed to be switched off at these asymptotic regions.
By differentiating with
respect to the source one generates matrix elements from $W[J]$ ,
\begin{equation}
     \frac{\delta W[J]}{\delta J(x)} = \frac{\langle 0,out|\phi
              (x)|in,0\rangle_J}
     {\langle 0,out|in,0\rangle_J} \equiv \bar{\phi} [J] .
\end{equation}
If we assume that the above expression can be reversed, the effective
action is defined as the Legendre transformation of the generating
functional,
\begin{equation}
     \Gamma[\bar{\phi}] = W[J] - J\bar{\phi} .
\end{equation}
This functional of $\bar\phi$ is the generator of the one-particle-irreducible
graphs (graphs that remain connected when any internal line is cut) and
contains all the quantum corrections to the classical action.
{}From (2.6) one may derive the dynamical equation for the effective mean field
$\bar{\phi}[0]$, $i.e.$ the matrix element of the field $\phi$ in the absence
of the source $J(x)$, as
\begin{equation}
   \left.  \frac{\delta\Gamma [\bar{\phi}]}{\delta
     \bar{\phi}}\right|_{\bar{\phi}=\bar{\phi}[0]} =0 ,
\end{equation}
which expresses the quantum corrections to the
classical equation as a variational problem of the effective action.

In order to work with expectation values rather than matrix elements one
can define a new generating functional whose dynamics is determined by two
different external classical sources $J_+$ and $J_-$, by letting the in vacuum
evolve independentenly under these sources,
\begin{equation}
     e^{iW[J_+,J_-]} = \sum_\alpha \langle 0,in|\alpha,T\rangle_{J_-}
             \langle\alpha ,T|in,0\rangle_{J_+} .
\end{equation}
Here we have assumed that $\{|\alpha,T\rangle\}$ is a complete basis of
eigenstates of the field operator $\phi (x)$ at some future time $T$, $i.e.$
$\phi(T,\mbox{\bf x})|\alpha,T\rangle=\alpha(\mbox{\bf x})|\alpha,T\rangle$.
Then (2.8) may be written according to (2.2) as
\begin{eqnarray}
         e^{iW[J_+,J_-]} \equiv \int d\alpha \langle 0,in|
                  T^{(a)}e^{-i\int_{-\infty}^Tdt\int d^{n-1}x
                  J_-(x)\phi(x)}|\alpha,T\rangle\nonumber \\ \times
                  \langle\alpha,T| T^{(t)}e^{i\int_{-\infty}^Tdt\int
                  d^{n-1}x J_+(x)\phi(x)}|in,0\rangle ,
\end{eqnarray}
where $T^{(t)}$ and $T^{(a)}$ mean, respectively, time and anti-time
ordered operators and $d\alpha$ means $d\alpha = \Pi_{\mbox{\bf x}}
d\alpha(\mbox{\bf x})$
where $\mbox{\bf x}$ are the points of the hypersurface
$\Sigma$ defined by $t=T$.
The generating functional has also a path integral representation,
\begin{equation}
      e^{iW[J_+,J_-]}=\int d\alpha\int {\cal D}[\phi_-]
           e^{-i\left(S[\phi_-] +J_-\phi_-\right)}
          \int{\cal D}[\phi_+] e^{i\left(S[\phi_+] +J_+\phi_+\right)},
\end{equation}
with the boundary conditions that $\phi_+=\phi_-=\alpha$ on $\Sigma$ and
that the fields $\phi_+$ and $\phi_-$ are pure negative and pure positive
modes, respectively, in the in region $i.e.$ $\phi_{\pm}\rightarrow
e^{\pm i\omega t}$ at $t\rightarrow -\infty$ (vacuum boundary
conditions in the remote past). In a more compact form one may write,
\begin{equation}
     e^{iW[J_+,J_-]} = \int {\cal D}[\phi_+]{\cal D}[\phi_-]
           e^{i\{S[\phi_+]
          +J_+\phi_+-S[\phi_-]-J_-\phi_-\}},
\end{equation}
where it is understood that the sum is over all fields $\phi_+$, $\phi_-$ with
negative and positive frequency modes, respectively, in the remote past but
which coincide at time $t=T$. These boundary conditions can be made explicit
by substituting $m^2$ by $m^2-i\epsilon$, where $m$ is the field mass, in
$S[\phi_+]$ and by substituting $m^2$ by $m^2+i\epsilon$ in $S[\phi_-]$;
the latter is also some
times indicated by writing $S^*[\phi_-]$ instead of $S[\phi_-]$ \cite{CH}.
This
integral can be thought of as the path sum of two different fields evolving in
two different time branches \cite{Haj}, one going forward in time in the
presence of $J_+$ from the in vacuum to a time $t=T$, and the other backward
in time in the presence of $J_-$ from the time $t=T$ to the in vacuum, with
the  constraint $\phi_+=\phi_-$ on $\Sigma$. Because of such a path integral
representation, this formalism is often called closed time path formalism.

The functional $W[J_+,J_-]$ generates expectation values of the
field rather than matrix elements. For instance, we have,
\begin{equation}
      \left.\frac{\delta W[J_+,J_-]}{\delta J_+}\right|_{J_\pm =J} =
            \langle 0,in|\phi(x)|in,0\rangle_J\equiv\bar\phi [J] ,
\end{equation}
instead of equation (2.5).
This functional generates not only the desired expectation values of time
ordered field operators but also the anti-time ordered
ones in the same footing
\begin{equation}
   \left. \frac{\delta e^{iW[J_+,J_-]}}{i\delta J_+(x_1)\cdots
      (-i)\delta J_-(y_1)\cdots}\right|_{J_\pm =J=0} =
              \langle 0,in|T^{(a)}(\phi (y_1)\cdots)
              T^{(t)}(\phi (x_1)\cdots)|in,0\rangle .
\end{equation}

In analogy with the in-out formalism the in-in effective action is
defined as the Legendre transform of the new generating functional as,
\begin{equation}
     \Gamma [\bar{\phi}_+,\bar{\phi}_-] = W[J_+,J_-] - J_+\bar{\phi}_+ +
        J_-\bar{\phi}_-  ,
\end{equation}
where the external sources are functionals of the fields $\bar{\phi}_+$
and $\bar{\phi} _-$, through the definitions
\begin{equation}
    \frac{\delta W[J_+,J_-]}{\delta J_\pm} \equiv \pm\bar{\phi}_\pm
          [J_+,J_-] ,
\end{equation}
which we assume can be reversed.

{}From the definitions (2.14) and (2.15) we get the equation for the
expectation
values $\bar\phi_{\pm}[J_+,J_-]_J$, $i.e.$
\begin{equation}
       {\delta\Gamma[\bar\phi_+,\bar\phi_-]\over
               \delta\bar\phi_{\pm}}=\mp J_{\pm},
\end{equation}
and by taking $J_{\pm}=0$ in (2.15) we recover the equation for
the vacuum expectation value of the field $\bar{\phi}[0]\equiv
\bar{\phi}_\pm[0,0]=\langle 0,in|\phi(x)|in,0\rangle $,
\begin{equation}
    \left. \frac{\delta \Gamma [\bar{\phi}_+,\bar{\phi}_-]}
       {\delta \bar{\phi}_+}\right|_{\bar{\phi}_\pm
       = \bar{\phi}_\pm[0,0]\equiv\bar{\phi}[0]} = 0.
\end{equation}
This equation does not follow from a simple variational principle in terms of
a single field $\bar\phi$: in the in-in action we have two fields $\bar\phi_+$
and $\bar\phi_-$ which are treated independently and only when the sources
have been eliminated they become the vacuum expectation value. Note also that
$\Gamma[\bar\phi,\bar\phi]=0$ as a consequence of (2.14) and of
the fact that $W[J,J]=0$, which follows from (2.8) and the usual normalization
for the states. Equation (2.17) is a dynamical field equation which admits an
initial value formulation: the
solution $\bar\phi [0]= \bar\phi_\pm [0,0]$ is real and causal, $i.e.$ the
solution at one spacetime point depends only on data on the past of that
point \cite{Jor86}.

For a free field theory, $i.e.$ a theory with a quadratic action,  we can
compute $W_0[J_+,J_-]$ from (2.11), which becomes now a Gaussian integration
for the two independent fields $\phi_+$ and $\phi_-$. The corresponding
propagators will be determined by the very particular boundary conditions of
this problem. In fact, let us assume that the free action for $\phi_+$ is
$S_0[\phi_+]=-\int d^nx {1\over 2}(\partial_{\mu}\phi_+ \partial^{\mu}\phi_+
+(m^2-i\epsilon )\phi_+^2)$ and that we have an analogous action
$S^*_0[\phi_-]$ for $\phi_-$, then
the classical field equations are,
\begin{equation}
    (\Box -m^2\pm i\epsilon )\phi_{\pm}^0(x)=-J_{\pm}(x).
\end{equation}
At this stage we can introduce the compact notation,
\begin{equation}
   \begin{array}{c}
       {\cal S}[\phi_a]= S[\phi_+]-S^*[\phi_-],
       \\  \\
       \phi_a(x)=\left(
              \begin{array}{c}
                  \phi_+ \\ \phi_-
              \end{array}\right),
       J_a(x)=\left(
              \begin{array}{c}
                  J_+ \\ -J_-
              \end{array}\right),
   \end{array}
\end{equation}
where $a$ and $b$ take the two values $+$ and $-$,
to simplify the mathematical expressions.
The solutions of the classical field equations, that satisfy the boundary
conditions, $\phi_{\pm}^0\rightarrow e^{\pm i\omega t}$, when $t\rightarrow
-\infty$ with $\omega \geq 0$, and
$\phi_+^0(T,\mbox{\bf x})= \phi_-^0(T,\mbox{\bf x}) $
in the hypersurface $\Sigma$, which we take
here at $t=T\rightarrow \infty$, can be written as,
\begin{equation}
         \phi^0_a(x)=-\int d^ny G^0_{ab}(x,y)J_b(y),
\end{equation}
where $G^0_{ab}(x,y)$ is the matrix
\begin{equation}
     G^0_{ab}= \left(\begin{array}{c}
                    \Delta_F\hspace{0.15in}-\Delta^+    \\
                    \Delta^-\hspace{0.15in}-\Delta_D
            \end{array} \right),
\end{equation}
defined with the Feynman, $\Delta_F$, Dyson, $\Delta_D$, and the
positive, $\Delta^+$, and negative, $\Delta^-$, Wightman functions:
\begin{eqnarray}
        \Delta_F(x-y)&=&-\int {d^np\over (2\pi)^n}
           {e^{ip\cdot (x-y)}\over p^2+m^2-i\epsilon},
           \nonumber \\
           \nonumber \\
        \Delta_D(x-y)&=&-\int {d^np\over (2\pi)^n}
           {e^{ip\cdot (x-y)}\over p^2+m^2+i\epsilon},
           \\
           \nonumber \\
        \Delta^{\pm}(x-y)&=&(\pm 2\pi i) \int {d^np\over (2\pi)^n}
           e^{ip\cdot (x-y)}\delta (p^2+m^2)\theta(\mp p^0).
           \nonumber
\end{eqnarray}
These propagators are solutions of the equations
\begin{equation}
       A^0_{ac}\, G^0_{cb}(x,y)=\delta^n(x-y)\delta_{ab},
\end{equation}
where the operator $A^0_{ab}$ is the diagonal matrix defined by
\begin{equation}
      A^0_{ab}=diag\left[\left(\Box -m^2+i\epsilon\right),
                  -\left(\Box -m^2-i\epsilon\right)\right].
\end{equation}
Furthermore the Feynman and Dyson Green functions have a mode decomposition
\begin{equation}
    \begin{array}{c}
      \Delta_F(x-y)= \theta(x^0-y^0)\Delta^-(x-y)-
                     \theta(y^0-x^0)\Delta^+(x-y),  \\ \\
      \Delta_D(x-y)= \theta(x^0-y^0)\Delta^+(x-y)-
                     \theta(y^0-x^0)\Delta^-(x-y),
    \end{array}
\end{equation}
which reflect the boundary conditions imposed over each classical field
solution $\phi^0_a(x)$, because the Green functions
$\Delta^\pm (x-y)\sim e^{\pm i\omega\cdot (x^0-y^0)}$
correspond to negative and positive frequency modes respectively
(here $\omega=\sqrt{\mbox{\bf p}^2+ m^2}$).
Note that it is also satisfied that $\partial_t\phi_+^0(T,\mbox{\bf x})=
\partial_t\phi_-^0(T,\mbox{\bf x})$ at $\Sigma$.
With these propagators to guarantee the boundary conditions the Gaussian
integration of (2.11) for a free field is,
\begin{equation}
      W_0[J_a]= -{1\over2}\int d^nx d^ny J_a(x)G_{ab}(x-y)J_b(y),
\end{equation}
where a term independent of $J_a$ has been discarded to satisfy $W_0[J,J]=0$,
and one can use now (2.13) to generate time ordered and anti-time ordered
expectation values of field operators. In particular we have that
\begin{equation}
   \begin{array}{l}
       \langle 0,in|T^{(t)}\phi(x)\phi(y)|in,0\rangle= i\Delta_F(x-y),
         \\ \\
       \langle 0,in|T^{(a)}\phi(x)\phi(y)|in,0\rangle= -i\Delta_D(x-y),
         \\ \\
       \langle 0,in|\phi(y)\phi(x)|in,0\rangle =-i\Delta^+(y-x)
         =i\Delta^-(x-y).
    \end{array}
\end{equation}

For interacting fields one can proceed as usual by writing
\begin{equation}
    e^{iW[J_a]}=
        e^{i\int d^nx {\cal L}_{int}({\delta\over i\delta J_a})}
        e^{iW_0[J_a]},
\end{equation}
where we have separated the Lagrangian into a free and an interacting part,
${\cal L}={\cal L}_0 +{\cal L}_{int}$.
Then one may continue in the usual perturbative fashion; however,
we are not going to consider self-interacting theories in this paper,
the only interaction will be with the gravitational field.

Let us now proceed to the main objective, namely, the evaluation of the
effective action $\Gamma[\bar\phi_a]$ up to the one loop order,
which corresponds to the first order expansion of $W[J_a]$ in powers of
$\hbar$. As usual \cite{AL}, if we assume that the action is bounded
from above then we can go to Euclidean space and solve (2.11)
by the steepest descent method; we keep, however,
the Minkowskian notation. Let us denote by
$\phi_+^{(0)}(x)$ and $\phi_-^{(0)}(x)$ the solutions of the classical field
equations which may, or may not, include self-interactions,
\begin{equation}
     {\delta {\cal S}[\phi_b^{(0)}]\over\delta\phi_a^{(0)}(x)}=-J_a(x),
\end{equation}
and let us expand the exponent in (2.11) about these background fields:
\begin{eqnarray}
     {\cal S}[\phi_a]+\int d^nx J_a(x)\phi_a(x)\hskip -0.3cm&=&
              \hskip -0.3cm{\cal S}[\phi^{(0)}_a]
              +\int d^nx J_a(x)\phi^{(0)}_a(x)
              \nonumber \\
           & &\hskip -2cm+{1\over2}\int d^nxd^ny
              \left[\phi_a(x)-\phi^{(0)}_a(x)\right]
              A_{ab}(x,y)
              \left[\phi_b(y)-\phi^{(0)}_b(y)\right]+\cdots
\end{eqnarray}
where
\begin{equation}
        A_{++}(x,y)\equiv
      \left({\delta^2S[\phi_+]\over\delta\phi_+(x)\delta\phi_+(y)}\right)_
         {\phi_+=\phi_+^{(0)}},
        A_{--}(x,y)\equiv -
      \left({\delta^2S^*[\phi_-]\over\delta\phi_-(x)\delta\phi_-(y)}\right)_
        {\phi_-=\phi_-^{(0)}},
\end{equation}
and, of course, $A_{+-}(x,y)=A_{-+}(x,y)\equiv  0$.
Substituting this into (2.11) the integration is now Gaussian and we can write
to this one loop order,
\begin{equation}
      e^{iW[J_a]}\simeq e^{iW^{(0)}[J_a]}\left(\det A_{ab}(x,y)\right)^{-1/2},
\end{equation}
where $W^{(0)}[J_a]={\cal S}[\phi^{(0)}_a]+\int d^nx J_a\phi^{(0)}_a$.
In terms of the propagator $G$, which is a
functional of the background fields $\phi_a^{(0)}(x)$ and takes a
$2\times2$ matrix form , $i.e.\, G(x,y)=A^{-1}(x,y)$, we can write (2.32) as
\begin{equation}
      W[J_a]\simeq W^{(0)}[J_a]- {i\over2}Tr(\ln G).
\end{equation}

The effective action, which is a functional of $\bar\phi_a$, can now be
explicitly found to the same order. Using (2.14), (2.15), and the fact
that $\bar\phi_a$ differs from $\phi_a^{(0)}$ by a term of order $\hbar$
we can show that $W^{(0)}[J_a]\simeq {\cal S}[\bar\phi_a]+
\int d^nx J_a\bar\phi_a $, so that finally we have
\begin{equation}
     \Gamma[\bar\phi_a]\simeq {\cal S}[\bar\phi_a]-{i\over2}Tr(\ln G).
\end{equation}

Now the equations for $\bar\phi_a$ can be deduced from (2.16) using the
explicit functional dependence on the fields given by (2.34). However we
should note from (2.17) that in order to get the field equations for the
expectation value of $\phi(x)$ we only need the explicit dependence of the
effective action on one of the fields, $\bar\phi_+$ or $\bar\phi_-$.
Therfore we are only interested in the dependence of (2.34) on
$\bar\phi_+$, say. Following Paz \cite{Paz} we can write
\begin{eqnarray}
     {\delta\over\delta\bar\phi_+(x)}\left( Tr (\ln G)\right)&=&
         -\int dy dz G_{ab}(z,y) {\delta\over\delta\bar\phi_+(x)}
         G_{ba}^{-1}(z,y)
         \nonumber \\
      &=&-\int dy dz G_{++}(z,y) {\delta\over\delta\bar\phi_+(x)}
         G_{++}^{-1}(z,y)
         \nonumber \\
      &=& {\delta\over\delta\bar\phi_+(x)}\left( Tr (\ln G_{++})\right),
\end{eqnarray}
where we have used that $G^{-1}_{ab}\equiv A_{ab}$
is diagonal, see (2.31). Thus we have,
\begin{equation}
      \Gamma [\bar\phi_+,\bar\phi_-]\simeq S[\bar\phi_+]
               -{i\over2}Tr (\ln G_{++}) +F
\end{equation}
where $F$ includes all the terms which do not contribute to the variation of
the field $\bar\phi_+$, $i.e. \left. {\delta F\over
\delta\bar\phi_+(x)}\right|_{\bar\phi_\pm =\bar\phi}=0$.
This expression
is very similar to the one loop in-out effective action which is
given by the above equation (2.36) where $G_{++}$
is substituted by the Feynman
propagator $\Delta_F$, the main difference is in the boundary conditions: the
propagator $G_{++}$ is defined as an expectation value and not as an
in-out matrix element.

This formalism can be extended to curved spacetimes without difficulties
assuming that the spacetime is globally hyperbolic \cite{Jor86}. The
hypersufaces of constant time are now Cauchy hypersurfaces and the in and
out states are defined in the Cauchy hypersurfaces corresponding to
the far past and far future respectively. Now the spacetime integrals
must be performed with the volume element $d^nx\sqrt{-g}$ where $g_{\mu\nu}$
is the spacetime metric. The above expressions (2.20), (2.23) and (2.36) are
still valid except that now the Feynman, Dyson and Wightman functions have a
different representation to that of (2.22). If the spacetime is asymptotically
flat in the in and out regions the previous boundary conditions for the
fields $\phi^0_{\pm}(x)$ will also apply; if not, in order to be able to
define physically
meaningful in and out vacua we must assume that we are still able to define
positive and negative frequency solutions in the asymptotic regions. This is
always possible, for instance, if the asymptotic regions admit approximate
timelike Killing fields. But, generally, in a curved spacetime the in and
out vacua are not equivalent. Jordan \cite{Jor86} has shown that for
quantum scalar fields in a curved spacetime the field equations are real and
causal up to the  two loop order and he has also checked the unitarity of the
formalism restricted to vacuum states.

Before ending this section let us rewrite (2.36) in a more convenient form for
us. In general the propagator $G_{++}$ cannot be found exactly and has to be
evaluated perturbatively. For instance, in the next
section we will take perturbations $h_{\mu\nu}(x)$ to a given
background metric and only the exact propagator corresponding to the
background is known. Thus we write

\begin{equation}
        A_{ab}= A^0_{ab}+ (V^{(1)}_{ab}+V^{(2)}_{ab}+...)
\end{equation}
where $A^0_{ab}$ is the unperturbed (diagonal) operator whose propagator,
$G_{ab}^0$, is known,
\begin{equation}
        A^0_{ac}G^0_{cb}=\delta_{ab},
\end{equation}
and the diagonal operators $V^{(1)}_{ab}+V^{(2)}_{ab}+\cdots $
contain the perturbative
terms (in the next section they will correspond to perturbations of
order $|h_{\mu\nu}|$ and $|h_{\mu\nu}|^2$ respectively). We can write
\begin{eqnarray}
     G_{ab}&=&G^0_{ab}-G^0_{ac}(V^{(1)}_{cd}+V^{(2)}_{cd}\cdots )G_{db}
              \nonumber \\
           &=&G^0_{ab}-G^0_{ac}V^{(1)}_{cd}G^0_{db}
              -G^0_{ac}V^{(2)}_{cd}G^0_{db}
              +G^0_{ac}V^{(1)}_{cd}G^0_{de}V^{(1)}_{ef}G^0_{fb}+\cdots ,
\end{eqnarray}
where the products are operator products. In particular,
\begin{equation}
      G_{++}= G^0_{++}-G^0_{+a}V^{(1)}_{ab}G^0_{b+}
              -G^0_{+a}V^{(2)}_{ab}G^0_{b+}
              +G^0_{+a}V^{(1)}_{ab}G^0_{bc}V^{(1)}_{cd}G^0_{d+}+\cdots ,
\end{equation}
expanding the logarithmic term in (2.36) and using that
$A^0_{++}G^0_{+-}=0$, we finally get,
\begin{eqnarray}
      \Gamma[\bar\phi_+,\bar\phi_-]&\simeq &
            S[\bar\phi_+]- {i\over2}Tr (\ln G^0_{++})+ F
            \nonumber \\
         & &+{i\over2}Tr\left(\atop\right.
             V_+^{(1)}G^0_{++}
            +V_+^{(2)}G^0_{++}
            -{1\over2}V_+^{(1)}G^0_{++}V_+^{(1)}G^0_{++}
            \nonumber \\
         & &\hskip 3cm +V_+^{(1)}G^0_{+-}V_-^{(1)}G^0_{-+} +
            \cdots \left.\atop\right).
\end{eqnarray}
where we have defined $V_+^{(i)}\equiv V_{++}^{(i)}$ and
$V_-^{(i)}\equiv -V_{--}^{(i)}$ following (2.31).
Note that if it were not for the last term which involves the propagator
$G^0_{+-}$ this expression for the in-in effective action would agree with the
in-out effective action which involves only one field $\phi(x)$; see, for
instance, Hartle and Hu \cite{HH}.
Therefore the term containing the propagator
$G^0_{+-}$ is the only new term that contributes to the field equation for
$\bar\phi_+(x)$. It can be seen \cite{Paz} that the effect of the
last term in (2.41) is to make the field equation for $\bar\phi(x)$
causal: if one takes the derivative of the in-in effective action with
respect to  $\bar\phi_+(x)$ and puts $\bar\phi_+=\bar\phi_-=\bar\phi$, the
resulting field equation is causal.

Notice that in the case of a free scalar field on a nearly flat background the
propagator $G^0_{ab}$ which solves (2.38), corresponding to the flat
background, and which provides the boundary conditions of the in-in problem
is simply given by $G^0_{++}=\Delta_F$,  $G^0_{--}=-\Delta_D$,
$G^0_{-+}=\Delta^-$ and $G^0_{+-}=-\Delta^+$, as can be seen from (2.23). In
fact, from (2.31) $A^0$ is, in this case, the operator defined in (2.24).

\section{In-in effective action}
\setcounter{equation}{0}

In this section we compute the in-in effective action (2.41) for a conformal
field  in a nearly conformally flat spacetime. The cosmological
background consists of a spatially flat homogeneous FRW with small
perturbations as,
\begin{equation}
     \tilde g_{\mu\nu}(x) \equiv a^2(\eta)(\eta_{\mu\nu}+h_{\mu\nu}(x))
            \enspace .
\end{equation}
where $a(\eta) = \mbox{\rm exp}(\omega (\eta))$ is the conformal factor,
$\eta$ is the conformal time $d\eta=dt/a$, $t$ is the cosmological time, and
$h_{\mu\nu}(x)$ is a symmetric tensor representing arbitrary small
perturbations; we take the metric signature $(-+\cdots +)$. The classical
action for a free ($i.e.$ with no self-interactions)  massless conformally
coupled scalar field $\Phi(x)$
is given by
\begin{equation}
   S_m[\tilde g_{\mu\nu},\phi] = - \frac{1}{2}\int d^nx\sqrt{-\tilde g}
      \left[\tilde g^{\mu\nu}\partial_\mu\Phi\partial_\nu\Phi +
      \xi(n)\tilde R\Phi^2\right] ,
\end{equation}
where $\xi(n)={n-2\over4(n-1)}$, $\tilde R$ is the Ricci scalar for the
metric $\tilde g_{\mu\nu}$, and we take the spacetime dimensions, $n$,
arbitrary for the moment in view of dimensional regularization. Because of
the conformal coupling one may simplify the problem by defining a new matter
field $\phi(x)$ and a new metric $g_{\mu\nu}(x)$ as,
\begin{equation}
 \phi(x)\equiv e^{{n-2\over2}\omega(\eta)}\Phi(x),\ \ \ g_{\mu\nu}(x)\equiv
    \eta_{\mu\nu}+ h_{\mu\nu}(x).
\end{equation}
Then the action (3.2), after integration by parts and assuming no
contributions
of the surface integrals, is equivalent to the action for the field $\phi(x)$
in the nearly flat metric $g_{\mu\nu}$,
\begin{equation}
   S_m[ g_{\mu\nu},\phi] = - \frac{1}{2}\int d^nx\sqrt{- g}
      \left[ g^{\mu\nu}\partial_\mu\phi\partial_\nu\phi +
      \xi(n) R\phi^2\right] ,
\end{equation}
where $R$ is the Ricci scalar for the metric $g_{\mu\nu}(x)$.
Therefore the problem has been reduced to that of a scalar field $\phi(x)$
in a nearly flat background. We must keep in mind that
the physical field is $\Phi(x)$, but the fact that the two fields differ
by just a power of the conformal factor, which is a function of time only,
simplifies considerable the connection between the vacua of the two fields.
For instance, a positive frequency mode in flat space will correspond
to a positive frequency mode in the conformally related space.
Since for a free field in flat space the in and out vacua are equivalent
(there is no particle creation) the same is true for the vacua of the
conformal field in the conformally flat background.
As a consequence non trivial quantum effects can
be produced only by the breaking of conformal flatness which in this case
is due to the coupling of the quantum field with the gravitational
perturbations. The above action can be expanded in terms of these
perturbations as,
\begin{equation}
     S_m[ h_{\mu\nu},\phi] = S^{(0)}_m[\phi] +\sum^\infty_{n=1}
        S^{(n)}_m[h_{\mu\nu},\phi] ,
\end{equation}
where the first term is simply the action for the field $\phi$ in flat
spacetime, and the higher perturbative terms carry all the information on the
interaction with the perturbations.

Since we are interested in deriving the semiclassical correction to Einstein's
equations due to the quantum effect of the scalar field but keeping the
gravitational field as classical, we have to add to the effective action
$\Gamma_m$ ($m$ stands for matter fields),
the classical action of the gravitational field
$S_g[\tilde g_{\mu\nu}]$. We should also add the action of any other
classical source but we shall ignore this for simplicity;
note that its effects on the semiclassical
equations may be taken into account by simply adding the corresponding
classical stress tensor to the quantum stress tensor. Furthermore, in order
to renormalize the effective action it is sufficient to add to the
usual Einstein's action, terms quadratic in the Riemann tensor,
\begin{eqnarray}
     S_g[\tilde g_{\mu\nu}]&\equiv & \int d^nx
        (-\tilde g(x))^{1/2} \left\{ \frac{1}{16\pi G_N}\tilde R(x)\right.
        \nonumber \\
     & &\hskip1cm
        +\frac{\mu^{n-4}}{2880\pi^2(n-4)}\left[ \tilde R_{\mu\nu\alpha\beta}
        (x)\tilde R^{\mu\nu\alpha\beta}(x)-\tilde R_{\mu\nu}(x)
        \tilde R^{\mu\nu}(x)\right]\left. {\atop }\right\},
\end{eqnarray}
where $\mu$ is an arbitrary mass scale which will be useful in dimensional
regularization. The quadratic terms with poles at $n=4$ are those which are
necessary to cancel the divergencies of $\Gamma_m$; notice that with this
election one obtains the correct trace anomaly.

Following the previous section we introduce now two fields $\phi_+(x)$ and
$\phi_-(x)$ which at some future hypersurface $\Sigma$  coincide:
$\phi_+(x)=\phi_-(x)$. Since the scalar field is proved by the gravitational
field we must assume that the two fields evolve in two different geometries
$g^+_{\mu\nu}(x)$ and  $g^-_{\mu\nu}(x)$ respectively where
$g^\pm_{\mu\nu}(x)\equiv\eta_{\mu\nu}+h^\pm_{\mu\nu}(x)$,
and since we assume that the
fields have no interaction other than the gravitational we need not introduce
the classical external currents $J_{\pm}(x)$. Thus we shall write the total
in-in effective action for the gravitational and the matter fields as,
\begin{equation}
     \Gamma_{(ii)} [\omega ,h^\pm _{\mu\nu}] =
            S_g [\omega ,h^+_{\mu\nu}]
           -S_g [\omega ,h^-_{\mu\nu}]
           +\Gamma_m [h^+_{\mu\nu},\phi_+; h^-_{\mu\nu},\phi_-],
\end{equation}
where  $\Gamma_m [h^+_{\mu\nu},\phi_+; h^-_{\mu\nu},\phi_-]$ contains the
quantum effects of the scalar field.

Following eq. (2.41) we write $\Gamma_m$ in a perturbative expansion in
$h_{\mu\nu}$ as $\Gamma_m=\Gamma_m^{(0)}+\Gamma_m^{(1)}+\Gamma_m^{(2)}+...$
and write only the terms which contribute to the variation of
$h^+_{\mu\nu}$, which are also those which contribute to the variation of
$\phi_+(x)$. Thus we can write,
$\Gamma^{(0)}_{(ii)}=S_g^{(0)}+\Gamma_m^{(0)}$,
$\Gamma^{(1)}_{(ii)}=S_g^{(1)}+\Gamma_m^{(1)}$ and
$\Gamma^{(2)}_{(ii)}=S_g^{(2)}+\Gamma_m^{(2)}$, where
\begin{eqnarray}
   \Gamma^{(0)}_m[\bar\phi_+]&=&
            S_m^{(0)}[\bar\phi_+]
            -{i\over2} Tr (\ln G^0_{++}),\nonumber \\
   \Gamma^{(1)}_m[h^+_{\mu\nu},\bar\phi_+]&=&
            S_m^{(1)}[\bar\phi_+]
            +{i\over2} Tr (V^{(1)}_+ G^0_{++}),\nonumber \\
   \Gamma^{(2)}_m[h^+_{\mu\nu},\bar\phi_+]&=&
            S_m^{(2)}[\bar\phi_+]+{i\over2} Tr (V^{(2)}_+G^0_{++})
            \nonumber \\ & &
            -{i\over4}Tr (V^{(1)}_+ G^0_{++}V^{(1)}_+ G^0_{++})
            +{i\over2}Tr (V^{(1)}_+ G^0_{+-}V^{(1)}_- G^0_{-+}).
\end{eqnarray}

To compute $V^{(1)}$ and $V^{(2)}$ we use (2.31) and (2.37) and expand
$S_m[\phi]$ as
in (3.5),
\begin{equation}
      S_m^{(0)}[\phi]= -{1\over2}\int d^nx
                   \left[\eta^{\mu\nu}\partial_\mu\phi
                   \partial_\nu\phi\right],
\end{equation}
\begin{equation}
      S_m^{(1)}[h_{\mu\nu},\phi]= {1\over2}\int d^nx
                  \left[\bar h^{\mu\nu}\partial_\mu\phi
                  \partial_\nu\phi-\xi(n)R^{(1)}\phi^2\right],
\end{equation}
\begin{equation}
      S_m^{(2)}[h_{\mu\nu},\phi]= -{1\over2}\int d^nx
                  \left[\hat h^{\mu\nu}\partial_\mu\phi
                  \partial_\nu\phi
                  +\xi(n)\left(R^{(2)}+{1\over2}hR^{(1)}\right)
                  \phi^2\right].
\end{equation}
where $\bar h_{\mu\nu}\equiv h_{\mu\nu}-{1\over2}h\eta_{\mu\nu}$,
$\hat h_{\mu\nu}\equiv {h_\mu}^\alpha h_{\alpha\nu}-{1\over2}
hh_{\mu\nu}+{1\over8}h^2\eta_{\mu\nu}-{1\over4}h_{\alpha\beta}
h^{\alpha\beta}\eta_{\mu\nu}$ and $R^{(1)}$ and $R^{(2)}$
are the first and second order terms, respectively, of the scalar
curvature (see equation B.9 from the appendix).
{}From these expressions one gets the operators $A^0$, $V^{(1)}$ and
$V^{(2)}$ by diffenciation with respect to the field $\phi(x)$,
\begin{equation}
      A^0\phi(x)=\Box\phi(x),
\end{equation}
\begin{equation}
      V^{(1)}(x)\phi(x)=-\left[
                \left(\partial_\mu \bar h^{\mu\nu}(x)\right)\partial_\nu
                +\bar h^{\mu\nu}(x)\partial_\mu\partial_\nu
                +\xi(n)R^{(1)}\right]\phi(x),
\end{equation}
\begin{equation}
       V^{(2)}(x)\phi(x)=\left[
                \left(\partial_\mu \hat h^{\mu\nu}(x)\right)\partial_\nu
                +\hat h^{\mu\nu}(x)\partial_\mu\partial_\nu
                -\xi(n)\left(R^{(2)}+{1\over2}hR^{(1)}\right)\right]\phi(x)
\end{equation}
The explicit form of the operator $V^{(2)}$ will not be needed, however.
Now we can write the propagator, from (3.12), as,
\begin{equation}
      G^{0}_{++}(x,x')= \Delta_F(x,x')= -\int
            {d^nk\over(2\pi)^n}{e^{ik\cdot(x-x')}\over k^2-i\epsilon},
\end{equation}
recall that this is the propagator for the field $\phi(x)$, the propagator
for the physical field $\Phi(x)$ is related to this by the conformal
factor \cite{BD}, but we do not need it here. All we need to know is that the
boundary conditions for the physical field are determined by the
boundary conditions of $\phi(x)$ in flat space.

The effective action depends on the fields $\bar\phi_\pm$ and the
metric perturbations $h^\pm_{\mu\nu}$. We may obtain the equation for the
field $\bar\phi (x)$ (field equation) by functional derivation with
respect to $\bar\phi_+$, $i.e.$ equation (2.17), and the equations for
the metric perturbations (backreaction equations) by functional derivation
with respect to $h^+_{\mu\nu}$. Our primary interest is to obtain
the backreaction equations. From equation (2.17) one can see that
the vacuum expectation value $\bar\phi\equiv\langle 0,in|
\phi |in,0\rangle = O(h_{\mu\nu})$ (it would be zero in flat space), and
thus the contribution to the backreaction equations coming from the
matter action term will be at least of second order in the
perturbations when the field equations are substituted.
Formally, one can compute the effective action as a
functional of $h^\pm_{\mu\nu}$ only, and thus, the terms involving the field
$\bar\phi_+$ ($S^{(i)}_m[\bar\phi_+]$) are not necessary.

We can now compute each
of the (divergent) terms (3.8). As it is well known the first term
$\Gamma^{(0)}_m$ is easily renormalized by adding a suitable
counterterm that cancels the divergencies which lead to the conformal anomaly
\cite{BD} \cite{FHH}, but, this term does not play any role in
the stress tensor of the field because it is independent of $h_{\mu\nu}$.

We can now go to the next term in (3.8), $\Gamma_m^{(1)}$, but this
formally divergent (tadpole) term has no contribution, since
$Tr (V^{(1)}_+ G^0_{++})$ involves $n$-dimensional integrals of the form
$1/k^2$, $k_\alpha /k^2$, and $k_\alpha k_\beta /k^2$ (where $k_\alpha$ is
the integration momentum variable) which are identically zero in dimensional
regularization \cite{Lei}. Therefore there is no term linear
in $h_{\mu\nu}$ in the effective action, $i.e.$ we have $\Gamma^{(1)}_m=0$.

The first non trivial quantum contributions
to the stress tensor coming from $\Gamma_m$ are quadratic in
$h_{\mu\nu}$ and we thus need to compute $\Gamma^{(2)}_m$. Here again the
second (tadpole) term in (3.8), $Tr(V^{(2)}_+G^0_{++})$,
gives no contribution in dimensional
regularization, since the typical
integrals are of the same type that those of
$Tr(V^{(1)}_+G^0_{++})$. For this reason
we do not need the explicit form of the operator $V^{(2)}$.

The problem is thus reduced to the evaluation of the third and fourth terms in
(3.8). As we have stressed in section 2 the third term also appears
in the evaluation of the in-out
effective action and the fourth term is typical
of the in-in contribution to this order.
Let us evaluate the third term, $i.e.$
$T_1\equiv -{i\over4}Tr(V^{(1)}_+G^0_{++}V^{(1)}_+G^0_{++})$
(recall that $G^0_{++}=\Delta_F$, with $m=0$),
\begin{eqnarray}
      T_1 \hskip-0.3cm &=& \hskip-0.3cm
              -{i\over4}\int d^nxd^nx'
              V^{(1)}_+(x)\Delta_F(x,x') V^{(1)}_+(x')\Delta_F(x',x)
              \nonumber \\
          &=& \hskip-0.3cm
              -{i\over4}\int d^nxd^nx'\int {d^np\over(2\pi)^n}
              {d^nq\over(2\pi)^n}
              \nonumber \\
          & & \hskip0.5cm
              \times\left[\left(\left(
              \partial_\mu \bar h^{\mu\nu}_+(x)\right)\partial_\nu
              +\bar h^{\mu\nu}_+(x)\partial_\mu\partial_\nu
              +\xi(n)R^{(1)}_+(x)\right)
              {e^{iq\cdot(x-x')}\over q^2-i\epsilon}\right]
              \nonumber \\
          & & \hskip 1.5cm
              \times\left[\left(\left(
              \partial_{\alpha '} \bar h^{\alpha\beta}_+(x')\right)
              \partial_{\beta '}
              +\bar h^{\alpha\beta}_+(x')\partial_{\alpha '}
              \partial_{\beta '}
              +\xi(n)R^{(1)}_+(x')\right)
              {e^{ip\cdot(x'-x)}\over p^2-i\epsilon}\right].
\end{eqnarray}
We now introduce the projector $P^{\mu\nu}=\eta^{\mu\nu}-p^\mu p^\nu /p^2$,
the symbol $\eta^{\mu\nu\alpha\beta}$, $\bar h^{\mu\nu}\equiv
\eta^{\mu\alpha\nu\beta}h_{\alpha\beta}$, change the $p$ integration by
$p'\equiv q-p$, rename $p'$ as $p$ again, and write $T_1$ as,
\begin{eqnarray}
      T_1\hskip-0.3cm&=&\hskip-0.3cm
            -i\int d^nxd^ny h^+_{\mu\nu}(x)h^+_{\alpha\beta}(y)
            \int {d^np\over(2\pi)^n} e^{ip\cdot(x-y)}
            \hat K^{\mu\nu\alpha\beta}(p),
            \nonumber \\
            & & \\
            \hat K^{\mu\nu\alpha\beta}(p)
            \hskip-0.3cm&=&\hskip-0.3cm
            {1\over4}\int{d^nq\over(2\pi)^n}
            {1\over(q^2-i\epsilon)\left[ (p-q)^2-i\epsilon\right]}
            \nonumber \\
            & &\hskip0.2cm
            \times\left(\eta^{\rho\mu\tau\nu}(q-p)_\rho q_\tau
            -\xi(n)p^2P^{\mu\nu}\right)
            \left(\eta^{\lambda\alpha\sigma\beta}(q-p)_\lambda q_\sigma
            -\xi(n)p^2P^{\alpha\beta}\right).\nonumber
\end{eqnarray}
The momentum integrals can be computed in the standard way, see appendix D,
and expanding around $n=4$ we get after a
rather long calculation,
\begin{eqnarray}
         \hat K^{\mu\nu\alpha\beta}(p)
            &=& {p^4I_1(p)\over 1440}\left[\atop\right.
                (3P^{\mu\beta}P^{\nu\alpha}-P^{\mu\nu}P^{\alpha\beta})
                \nonumber \\
            & & \hskip1.5cm
                +{(n-4)\over 15}\left(
                8(P^{\mu\nu}P^{\alpha\beta} -3P^{\mu\beta}P^{\nu\alpha})+
                5P^{\mu\nu}P^{\alpha\beta}\right)
                \nonumber \\
            & & \hskip1.5cm
                +O(n-4)^2\left.\atop\right]
\end{eqnarray}
where
\begin{equation}
         I_1(p)= -{i\over8\pi^2}\left({1\over n-4}+{1\over2}
         \ln \left[ {p^2-i\epsilon\over\mu ^2_0} \right]+O(n-4)\right),
\end{equation}
and the parameter $\mu_0$ is a fixed parameter involving Euler's constant
$\gamma $. Now using the expressions of the Riemann components
in terms of the projectors $P_{\mu\nu}$ of appendix B.2 we can write,
\begin{eqnarray}
          T_1 \hskip-0.3cm&=& \hskip-0.3cm
                -{\alpha\over4}\left\{ {1\over n-4}\right.
                \int d^4x\left(3R^+_{\mu\nu\alpha\beta}(x)
                R^{+\mu\nu\alpha\beta}(x)-R^{+2}(x)\right)
                +{1\over3} \int d^4xR^{+2}(x)
                \nonumber \\
            & & \hskip0.8cm -\int d^4xd^4y\left[
                3R^+_{\mu\nu\alpha\beta}(x)R^{+\mu\nu\alpha\beta}(y)
                -R^+(x)R^+(y)\right]
                \mbox{\rm K}_1(x-y;\mu_0)
                \nonumber \\
            & & \hskip0.8cm +O(n-4)\left.\atop\right\},
\end{eqnarray}
where
\begin{equation}
   \mbox{\rm K}_1(x-y;\mu_0)\equiv
          -{1\over2}\int {d^4p\over(2\pi)^4} e^{ip\cdot(x-y)}
          \ln\left[ {(p^2-i\epsilon)\over\mu_0^2}\right],
\end{equation}
and $\alpha\equiv (2880\pi^2)^{-1}$.
Note that the divegent terms with a pole at $n=4$, are local and quadratic in
the curvature. They may be compensated by counterterms in the gravitational
part of the action $S_g^{(2)}$ coming from (3.6). Recall that the curvature
terms here depend on the metric $g_{\mu\nu}$ rather than the physical metric
$\tilde g_{\mu\nu}$.

Let us compute now the fourth term of $\Gamma_m^{(2)}$, $i.e.$
$T_2\equiv {i\over2}Tr(V^{(1)}_+G^0_{+-}V^{(1)}_-G^0_{-+})$,
which depends on
the propagators $G^0_{+-}=-\Delta^+$ and $G^0_{-+}=\Delta^-$,
represented in (2.22) with $m=0$. We can write
\begin{eqnarray}
      T_2\hskip-0.3cm&=&\hskip-0.3cm
            -{i\over2}\int d^nxd^nx'
            V^{(1)}_+(x)\Delta^+(x,x')
            V^{(1)}_-(x')\Delta^-(x',x)
            \nonumber \\
        &=& \hskip-0.3cm
            (-2i\pi^2)\int d^nxd^nx'
            \int{d^np\over(2\pi)^n}{d^nq\over(2\pi)^n}
            \delta(q^2)\theta(-q^0) \delta(p^2)\theta(p^0)
            \nonumber \\
        & & \hskip 0.5cm
            \times\left[\left(
            \partial_\mu \bar h^{\mu\nu}_+(x)\right)\partial_\nu
            +\bar h^{\mu\nu}_+(x)\partial_\mu\partial_\nu
            +\xi(n)R^{(1)}_+(x)\right]
            e^{iq\cdot(x-x')}
            \nonumber \\
        & & \hskip 1.5cm
            \times\left[\left(
            \partial^{'}_\alpha \bar h^{\alpha\beta}_-(x')\right)
            \partial^{'}_\beta
            +\bar h^{\alpha\beta}_-(x')\partial^{'}_\alpha
            \partial^{'}_\beta
            +\xi(n)R^{(1)}_-(x')\right]
            e^{ip\cdot(x'-x)},
\end{eqnarray}
changing the integration variable from $p$ to $q-p$ as in the previous
case we can write,
\begin{eqnarray}
       T_2\hskip-0.3cm&=&\hskip-0.3cm
          -8i\pi^2\int d^nxd^ny h^+_{\mu\nu}(x)h^-_{\alpha\beta}(y)
          \int {d^np\over(2\pi)^n} e^{ip\cdot(x-y)}
          \hat L^{\mu\nu\alpha\beta}(p),
          \nonumber \\
          & & \\
          \hat L^{\mu\nu\alpha\beta}(p)
          \hskip-0.3cm&=&\hskip-0.3cm
          {1\over4}\int{d^nq\over(2\pi)^n}
          \delta((p-q)^2)\theta(q^0-p^0)\delta(q^2)\theta(-q^0)
          \nonumber \\
          & &\hskip0.2cm
          \times\left(\eta^{\rho\mu\tau\nu}(q-p)_\rho q_\tau
          -\xi(n)p^2P^{\mu\nu}\right)
          \left(\eta^{\lambda\alpha\sigma\beta}(q-p)_\lambda q_\sigma
          -\xi(n)p^2P^{\alpha\beta}\right).\nonumber
\end{eqnarray}
After performing the phase space integrals and expanding around $n=4$, see
appendix D, we obtain
\begin{equation}
      \hat L^{\mu\nu\alpha\beta}(p)=
               {p^4I_2(p)\over 1440}\left[ (3P^{\mu\beta}P^{\nu\alpha}-
               P^{\mu\nu}P^{\alpha\beta}) + O(n-4)\right],
\end{equation}
where
\begin{equation}
      I_2(p)={1\over8\pi^2}\left[ {1\over 4\pi}
             \theta (-p^2)\theta (-p^0)+O(n-4)\right],
\end{equation}
which has no poles at $n=4$, therefore the term $T_2$ requires no
counterterms in the action to be renormalized. Using the
expressions of the Riemann components in terms of $P_{\mu\nu}$
(see appendix B.2) we get,
\begin{equation}
      T_2 =
          {\alpha\over2} \int d^4xd^4y
          \left[3R^+_{\mu\nu\alpha\beta}(x) R^{-\mu\nu\alpha\beta}(y)
          -R^+(x)R^-(y)\right]
          \mbox{\rm K}_2(x-y)
          +O(n-4),
\end{equation}
where
\begin{equation}
      \mbox{\rm K}_2(x-y)\equiv -{1\over 2}\int {d^4p\over(2\pi)^4}
               e^{ip\cdot(x-y)}(2\pi i)\theta(-p^2)\theta(-p^0).
\end{equation}
Here again the Riemann components refer to the metric $g_{\mu\nu}$. We now
have $\Gamma_m^{(2)}= T_1+T_2$ which must be renormalized by adding the
gravitational action up to the second order in $h_{\mu\nu}$. The explicit
expansion of $S_g$, in (3.6), up to this order in terms of the curvature
components of the metric $g_{\mu\nu}$ is:
\begin{eqnarray}
   S_g[\tilde g_{\mu\nu}] \hskip-0.3cm &\equiv & \hskip-0.3cm
                 S_g^{(0)}+S_g^{(1)}+ S_g^{(2)}+...
                 \nonumber \\ \hskip-0.3cm
              &=&\hskip-0.3cm
                 {1\over16\pi G_N}\int d^4x (-g(x))^{1/2}e^{2\omega}
                 [R(x)+6\omega_{;\mu} \omega^{;\mu}]
                 \nonumber \\
              & &\hskip-0.7cm
                 +{\alpha\over 4(n-4)}\int d^4x
                 [3R_{\mu\nu\alpha\beta}(x) R^{\mu\nu\alpha\beta}(x)-R^2(x)]
                 \nonumber \\
              & &\hskip-0.7cm
                 +\alpha\int d^4x
                 [R_{\mu\nu\alpha\beta}(x) R^{\mu\nu\alpha\beta}(x)
                 -R_{\mu\nu}(x)R^{\mu\nu}(x)]\ln(\mu e^\omega)
                 \nonumber \\
              & &\hskip-0.7cm
                 +\alpha\int d^4x(-g(x))^{1/2}
                 [2R^{\mu\nu}\omega_{;\mu}\omega_{;\nu}
                 + R\Box_g\omega
                 -4(\Box_g\omega)\omega_{;\nu}\omega^{;\nu}
                 -3(\Box_g\omega)^2
                 -2(\omega_{;\nu}\omega^{;\nu})^2]
                 \nonumber \\
              & &\hskip-0.7cm
                 +O(n-4),
\end{eqnarray}
where we have dropped the $+$ sign on
the fields for simplicity. Finally adding
these terms to $\Gamma_m$ and including only the terms which contribute to the
variation of $h^+_{\mu\nu}$ we get the renormalized effective action,
\begin{eqnarray}
        \Gamma ^{(2)}_{(ii)}
     &\hskip -0.4 cm [& \hskip -0.45cm\omega,h^\pm_{\mu\nu}]
        =\int d^4x(-\tilde g^+(x))^{1/2}\left[ \frac{\tilde R^+(x)}{16\pi G_N}
        -{\alpha\over 12} \tilde R^+(x)\tilde R^+(x)\right]
        \nonumber \\
     & &+2\alpha \int d^4x(-g^+(x))^{1/2}\left[  G^{+\mu\nu}(x)
        \omega_{;\mu}\omega_{;\nu} + \Box_g\omega(\omega_{;\nu}\omega^{;\nu})
        +{1\over 2}(\omega_{;\mu} \omega^{;\mu})^2\right]
        \nonumber \\
     & &+\alpha\int d^4x(-g^+(x))^{1/2}\left[ (R^+_{\mu\nu\alpha\beta}
        (x) R^{+\mu\nu\alpha\beta}(x)- R^+_{\mu\nu}(x)R^{+\mu\nu}(x))
        \right] \omega (x)  \nonumber \\
     & &+{\alpha\over 4}\int d^4x d^4y (-g^+(x))^{1/2}(-g^+(y))^{1/2}
        \nonumber \\
     & &\hskip 2cm\times
        \left[ 3R^+_{\mu\nu\alpha\beta}(x) R^{+\mu\nu\alpha\beta}(y)
        -R^+(x)R^+(y) \right]\mbox{\rm K}_1(x-y;\bar\mu) \nonumber \\
     & &+{\alpha\over 2} \int d^4x d^4y (-g^+(x))^{1/2}(-g^-(y))^{1/2}
        \nonumber \\
     & &\hskip 2cm\times
        \left[ 3R^+_{\mu\nu\alpha\beta}(x) R^{-\mu\nu\alpha\beta}(y)
        -R^+(x)R^-(y) \right]\mbox{\rm K}_2(x-y)\nonumber \\
     & &+{\rm O}(h^3_{\mu\nu}),
\end{eqnarray}
where $\bar\mu\equiv\mu\mu_0$ and we have substituted $1$ by the
volume densities
$\sqrt{-g}$ in all the integrals involving quadratic curvature terms of the
metric $g_{\mu\nu}$ in order to facilitate the identification of the exact
variational formulae of the appendix E needed in the computations of
the next section.

If one is interested
in the production of particles, the in-in effective action is not suitable
because the probability of particle creation is related to the transition
amplitude from the in to the out vacua, and this amplitude is directely
related to the in-out effective action \cite{Har}. Let us write the
vacuum persistence amplitude in the presence of an external source, $i.e.$
$\langle 0,out|in,0\rangle_J=\exp (iW[J])$.
The probability of pair creation is
proportional to the imaginary part of $W$
$$P=2 Im W,$$
but if we take $J=0$ and consider the quantum fields propagating in the
gravitational background, then $W$ is just the in-out effective action:
$\Gamma_{(io)}$. The calculations leading to such an action are similar
to those for the in-in case, although they are simpler because
we do not need to introduce two fields. The main work has already been done:
consider just a single field in (3.8) and ignore the term $T_2$. The
renormalized action is obtained again
by adding the gravitational action (3.28),
the final result can be read directly from (3.29): ignore the plus indices
and the term involving $\mbox{\rm K}_2(x-y)$.
The non local term now includes $\mbox{\rm K}_1(x-y;\bar\mu)$ only, this
term is complex and it is responsible for the particle creation effect. It
turns out that the pair creation probability is given, however, by a local
term as is well known \cite{Har}\cite{inh}. In fact, the
imaginary part of the kernel $\mbox{\rm K}_1(x-y;\bar\mu)$ is,
\begin{equation}
       Im \mbox{\rm K}_1(x-y;\bar\mu)={\pi\over 2}\int {d^4p\over(2\pi)^4}
            e^{ip\cdot(x-y)}\theta (-p^2).
\end{equation}
{}From the expression for $\Gamma_{(io)}$ and (3.30),
after performing the $x$ and $y$ integrations which lead to the
Fourier transform of the curvature tensor, $R_{\mu\nu\alpha\beta}(p)$,
we get
\begin{equation}
         P=2Im \Gamma_{(io)}={\alpha\pi\over 4}\int {d^4p\over (2\pi )^4}
             \left[3R_{\mu\nu\alpha\beta}(p)
             R^{\mu\nu\alpha\beta}(-p)-R(p)R(-p)\right]\theta (-p^2).
             \nonumber
\end{equation}
Finally, using that the Gauss-Bonnet
topological invariant is zero, the above relation can be written in terms
of the Fourier transform of the Weyl tensor of the physical metric
$\tilde g_{\mu\nu}$ as,
\begin{equation}
         P={1\over 960\pi }\int {d^4p\over (2\pi )^4}
               |\tilde C_{\mu\nu\alpha\beta}(p)|^2\theta (-p^2)\theta (-p^0),
\end{equation}
in agreement with the expressions computed by other means \cite{inh}.

\section{Semiclassical equations}
\setcounter{equation}{0}

In this section we obtain the quantum mechanically corrected Einstein's
equations due to the presence of a massless conformal scalar quantum field.
The semiclassical equations for the metric perturbation can be found by
functional differentiation of the in-in effective action (3.29) with respect
to $h^+_{\mu\nu}(x)$ and then restricting $h^+_{\mu\nu}(x)
=h^-_{\mu\nu}(x)=h_{\mu\nu}(x)$, as
\begin{equation}
     \left.{\delta\Gamma^{(2)}_{(ii)} [\omega , h^\pm _{\mu\nu}]\over
     \delta h^+_{\mu\nu}(x)}
     \right|_{h^\pm_{\mu\nu}=h_{\mu\nu}}=0 .
\end{equation}

{}From equation (4.1) it is easy to derive the equations of
motion to first order; we use that for an arbitrary functional
$A[\tilde g_{\mu\nu}]$,
\begin{equation}
     \frac{\delta A [\omega , g_{\mu\nu}]} {\sqrt{-g}\delta g_{\mu\nu}}=
     e^{6\omega}
     \frac{\delta A [\tilde g_{\mu\nu}]}
     {\sqrt{-\tilde g}\delta \tilde g_{\mu\nu}},
\end{equation}
to find the variation of the first two terms of (3.29).
Notice that  we assume
that $\omega (x)$ is a scalar function independent of the metric (in general
we will assume that  $\omega (x)$ depends on the spacetime point only,
in particular, in  the flat FRW case it will be a function of the
cosmological time $t$ only).

Using the expressions listed in the appendix E one can show that the
semiclassical equations can be written as
\begin{eqnarray}
      & &\hskip-1cm
         e^{6\omega}\left[ -{1\over 16\pi G_N}\left(\tilde G^{\mu\nu}_{(0)}
                            +\tilde G^{\mu\nu}_{(1)}\right)
                            -{\alpha\over 12}\left(\tilde B^{\mu\nu}_{(0)}
                            +\tilde B^{\mu\nu}_{(1)}\right)
                            +{\alpha\over 2}\left(\tilde H^{\mu\nu}_{(0)}
                            +\tilde H^{\mu\nu}_{(1)}\right)
                     \right] \nonumber \\
     & &\hskip-1cm
        -\alpha\tilde R^{(0)}_{\alpha\beta} C^{\mu\alpha\nu\beta}_{(1)}
                            +{3\alpha\over 2} \left[
                            -4(C^{\mu\alpha\nu\beta}_{(1)}\omega )
                            _{,\alpha\beta}
 			    +\int d^4y A^{\mu\nu}_{(1)}(y)
                            \mbox{\rm H}(x-y;\bar\mu)
                            \right]+ O(h^2_{\mu\nu})= 0,
\end{eqnarray}
where $G^{\mu\nu}(x)$ is the Einstein's tensor, $C^{\mu\alpha\nu\beta}(x)$
the Weyl's tensor, $B^{\mu\nu}(x)$ and $A^{\mu\nu}(x)$ are the exact spacetime
tensors given by the variation of $\int d^4x R^2(x)$ and
$\int d^4x C_{\mu\alpha\nu\beta}C^{\mu\alpha\nu\beta}$, respectively
(see appendix E.2 and use that $C_{\mu\alpha\nu\beta}C^{\mu\alpha\nu\beta}=
R_{\mu\alpha\nu\beta}R^{\mu\alpha\nu\beta}-2R_{\mu\nu}R^{\mu\nu}
+{1\over3}R^2$), with respect to an arbitrary metric $g_{\mu\nu}(x)$
$$
  B^{\mu\nu}(x)\equiv {1\over2}g^{\mu\nu}R^2
                      -2RR^{\mu\nu}
                      +2R^{;\mu\nu}
                      -2g^{\mu\nu}\Box_g R,
$$
$$
  A^{\mu\nu}(x)\equiv {1\over2}g^{\mu\nu}C_{\alpha\beta\rho\sigma}
                      C^{\alpha\beta\rho\sigma}
                      -R^{\mu\alpha\beta\rho}{R^\nu}_{\alpha\beta\rho}
                      +4R^{\mu\alpha}{R_\alpha}^\nu
$$
$$
         \hskip 1.5cm -{2\over3}RR^{\mu\nu}
                      -2\Box_g R^{\mu\nu}
                      +{2\over3}R^{;\mu\nu}
                      +{1\over3}g^{\mu\nu}\Box_g R,
$$
and $H^{\mu\nu}(x)$ is the spacetime tensor defined by
$$
 H^{\mu\nu}(x)\equiv -R^{\mu\alpha}{R_\alpha}^\nu
                     +{2\over3}RR^{\mu\nu}
                     +{1\over2}g^{\mu\nu}R_{\alpha\beta}R^{\alpha\beta}
                     -{1\over4}g^{\mu\nu}R^2.
$$
In equations (4.3) we are only interested in the expressions of these
tensors up to first order in the perturbations, as indicated by the
bracketed subindices and we recall that an over tilde on the tensors
refer to the physical metric.
The non-local part $\mbox{\rm H}(x-y;\bar\mu)$ is the
sum of the integrals (3.21) and (3.27),
\begin{eqnarray}
        \mbox{\rm H}(x-y;\bar\mu)
        \hskip-0.3cm&\equiv&\hskip-0.3cm
        \mbox{\rm K}_1(x-y;\bar\mu)+\mbox{\rm K}_2(x-y)
        \nonumber \\
        \hskip-0.3cm&=&\hskip-0.3cm
        -{1\over2}\int {d^4p\over(2\pi)^4}
        e^{ip\cdot(x-y)}
        \left\{\ln\left [ {(p^2-i\epsilon)\over\bar{\mu}^2}\right ]
        +(2\pi i)\theta(-p^2)\theta(-p^0)\right\},
\end{eqnarray}
which can be simplified to
\begin{equation}
       \mbox{\rm H}(x-y;\bar\mu) = -{1\over2}\int {d^4p\over(2\pi)^4}
       e^{ip\cdot(x-y)}
       \left\{\ln\left[{|p^2|\over\bar{\mu}^2}\right]
       +i\pi \theta(-p^2)sign(-p^0)\right\},
\end{equation}
by using $\ln (\pm i)=\pm i {\pi\over 2}$ and $\lim _{\epsilon\rightarrow 0}
\ln (\epsilon +ix)=\ln |x| + sign(x)i{\pi\over 2}$.
Note that this equation is real, in spite of appearences,
because the imaginary part of the integrand is an odd term with respect to
the integration variable $p^{\mu}$. One can also notice that
$\mbox{\rm H}(x-y;\bar\mu)$ differs from that
defined by Horowitz \cite{Hor80} by a factor $(1/4\pi )$.

Equations (4.3) are dynamical equations with higher order derivative
terms. When the background is flat they reduce to the field equations
studied by Horowitz \cite{Hor80} and Jordan \cite {Jor87}. As we have noted
in the previous section one can add a classical stress matter source to
these field equations.

To compare the functional method used in this paper with other techniques
and, in particular, to compare equations (3.29) and (4.3)
with previous results one can give, for exemple the energy-momentum
tensor of the quantum field  and the expression of the semiclassical
equations in two dimensions.

\subsection{Stress tensor to first order}

{}From (4.3) one can read the zero and first order vacuum expectation
value of the energy-momentum tensor of the scalar field,
\begin{eqnarray}
   \langle T^{\mu\nu}_{(0)} \rangle
          &=& \alpha\left[\tilde H^{\mu\nu}_{(0)}
              -{1\over 6} \tilde B^{\mu\nu}_{(0)}\right] \\
   \langle T^{\mu\nu}_{(1)} \rangle
           &=& \alpha\left[ (\tilde H^{\mu\nu}_{(1)}
              -2\tilde R^{(0)}_{\alpha\beta}
              \tilde C^{\mu\alpha\nu\beta}_{(1)}) -{1\over 6}
              \tilde B^{\mu\nu}_{(1)}\right. \nonumber \\
           & &\hskip1cm
              +3e^{-6\omega}\left.\left(
              -4( C^{\mu\alpha\nu\beta}_{(1)}\omega )_{,\alpha\beta}
              +\int d^4y A^{\mu\nu}_{(1)}(y)\mbox{\rm H}(x-y;\bar\mu)
              \right)\right]
\end{eqnarray}
The stress tensor to first order in $h_{\mu\nu}$, $\langle
T^{\mu\nu}_{(1)} \rangle$, is in agreement with that obtained
by Horowitz and Wald \cite{HW} and Starobinsky \cite{Sta81}.
On the other hand, the zeroth order tensor
$\langle T^{\mu\nu}_{(0)} \rangle$, which
gives the exact stress tensor for a conformal
scalar field in a conformally flat spacetime, agrees with that found by other
techniques \cite{DFCB} \cite{BD}. Note also that we recover the
trace anomaly result to this order in $h_{\mu\nu}$,
\begin{eqnarray}
    \langle {T^\mu}_{\mu} \rangle
                   &=& \langle T^{\ \mu}_{(0)\mu} \rangle
                       +\langle T^{\ \mu}_{(1)\mu} \rangle
                       +O(h^2_{\mu\nu}) \nonumber \\
                   &=& \alpha\left[\Box_{\tilde g}\tilde R
                       +\left( \tilde R^{\mu\nu}\tilde R_{\mu\nu}
                       -{1\over 3}\tilde R^2\right)\right]
                       +O(h^2_{\mu\nu}).
\end{eqnarray}

\subsection{Two dimensional gravity}

As an exercise we will derive here the semiclassical corrections to the stress
tensor of the scalar field in two spacetime dimensions. At the classical
level there is no dynamics for the gravitational field in two dimensions
because the Einstein tensor is identically zero, but the semiclassical
corrections due to the presence of a quantum field lead to non trivial
effects. Since we have worked in $n$ dimensions in order to
use dimensional regularization, we can use most of the work done in the
previous section to analize
the two dimensional case. From equations (3.17) and appendix D
it is not difficult to show that,
\begin{equation}
      \hat K^{(2D)}_{\mu\nu\alpha\beta}(q)={I^{(2D)}_1(q)\over 96}(n-2)q^4
      P_{\mu\nu}P_{\alpha\beta}+O(n-2)^2
\end{equation}
where
\begin{equation}
      I^{(2D)}_1(q)={1\over\pi (q^2-i\epsilon)}\left[{1\over n-2}
              +{\gamma\over 2}+O(n-2) \right] .
\end{equation}
Then, the in-out contribution to the effective action $T_1$ becomes,
\begin{equation}
      T_1=-{1\over 96\pi}\int d^2xd^2yR(x)\Delta^{(2D)}_F(x-y)R(y) ,
\end{equation}
where $\Delta^{(2D)}_F(x-y)$ is the Feynman propagator in two dimensions,
\begin{equation}
      \Delta^{(2D)}_F(x-y)=-\int {d^2q\over (2\pi )^2}{e^{iq\cdot (x-y)}
      \over q^2-i\epsilon} .
\end{equation}
{}From equation (3.23) it is straighforward to see that there is no
contribution to the effective action due to the typical in-in term $T_2$.
Quantum corrections appear to be non-local and quadratic in the
scalar curvature, but contrary to the four dimensional case there are
no divergent terms in the regularization process and counteraction
terms are not needed.
Finally one can express the effective action to one loop as,
\begin{eqnarray}
      \Gamma ^{(2D)}[\omega,h_{\mu\nu}]
    &=&{1\over 16\pi G_N}\int d^2x(-g(x))^{1/2} [R-2\Box_g\omega ]
       \nonumber \\
    &-&{1\over 96\pi }\int d^2xd^2y(-g(x))^{1/2}(-g(y))^{1/2}
       R(x)\Delta^{(2D)}_F(x-y)R(y)
       \nonumber \\
    &+& O(h^3_{\mu\nu}) .
\end{eqnarray}

The stress tensor of the quantum field can be obtained from the second term
of the above equation by differentiating with respect to the metric.
To order zero in the perturbation the energy-momentum tensor vanishes,
but to first order there is a non-local contribution,
\begin{equation}
      \langle T^{\mu\nu}_{(1)}(x) \rangle^{(2D)}=
          {1\over 24\pi}\int d^2y[\eta ^{\mu\nu}
          R_{,\alpha}^{\;\;\;\alpha}-R^{,\mu\nu}]
          \Delta^{(2D)}_F(x-y) + O(h^2_{\mu\nu}),
\end{equation}
and the trace is local $\langle T^{\,\mu}_{(1)\mu}(x)
\rangle^{(2D)}=R/ (24\pi ) + O(h^2_{\mu\nu})$ as
expected \cite{BD}. Note that in two dimensions there is no difference
between the in-out and the in-in effective actions, thus the use of the
usual in-out effective action to derive the semiclassical equations is
justified.

\section{Backreaction on the field of a static cosmic string}
\setcounter{equation}{0}

In this section, as a simple example,  we discuss the backreaction due
to one loop quantum effects on the gravitational field of a static cosmic
string. First, we must compute the vacuum expectation value of the stress
tensor of a conformally coupled massless scalar quantum field outside the
core of a straight and static cosmic string.

In the weak field approximation the
metric of a cosmic string can be seen as a small perturbation about
flat space,
\begin{equation}
      g_{\mu\nu}(x)=\eta_{\mu\nu}+h_{\mu\nu}(x).
\end{equation}
Let $T^{\mu\nu}_c$ be the stress tensor of a cosmic string (or of any other
classical source), the semiclassical equations up to first order in the
perturbation, can be written us,
\begin{equation}
     G^{\mu\nu}(x)=8\pi G_N\left[ T^{\mu\nu}_c(x)
                    + \langle T^{\mu\nu}_{(1)}(x)\rangle
                    +O(h^2_{\mu\nu})  \right] .
\end{equation}
There is no zero order correction, $\langle T^{\mu\nu}_{(0)} \rangle $,
to the classical Einstein's equations because the background is flat;
see (4.6). The vacuum expectation value
$\langle T^{\mu\nu}_{(1)} \rangle$ is obtained from equation
(4.7) when the conformal function is  $\omega =0$. We have
\begin{equation}
     \langle T^{\mu\nu}_{(1)}(x) \rangle = -{\alpha\over 6}B^{\mu\nu}_{(1)}(x)
                +3\alpha\int d^4y\mbox{\rm H}(x-y;\bar\mu)A^{\mu\nu}_{(1)}(y).
\end{equation}
where
\begin{eqnarray}
         B^{\mu\nu}_{(1)}(x)
     &=&  2\eta^{\mu\nu} {{{G_\alpha}^\alpha}_{,\beta}}^\beta
          -2 {{G_\alpha}^\alpha}^{,\mu\nu}, \nonumber \\
         A^{\mu\nu}_{(1)}(x)
     &=&  -2{{G^{\mu\nu}}_{ ,\alpha}}^{\alpha}
          -{2\over 3} {{G_\alpha}^\alpha}^{,\mu\nu}
          +{2\over 3}\eta ^{\mu\nu}{{{G_\alpha}^\alpha}_{,\beta}}^\beta .
\end{eqnarray}
Note that
the quantum correction term $\langle T^{\mu\nu}_{(1)} \rangle$ depends on
the Einstein tensor $G^{\mu\nu}$, thus one may use Einstein's equations to the
classical order, which is already first order in $h_{\mu\nu}$, to
substitute $G^{\mu\nu}$ by $8\pi G_N T^{\mu\nu}_c$.
This simplifies considerably the problem since the explicit
gravitational field of the string (or the classical source)
is not required to compute $\langle T^{\mu\nu}_{(1)} \rangle$.

The stress tensor of a static cosmic string along the $z$-axis can be
written in the thin line approximation as,
\begin{equation}
     {T_{c\mu}}^\nu (x,y)=-\mu\delta (x)\delta (y){\it diag}(1,0,0,1),
\end{equation}
where $\mu $ is the mass per unit lenght of the string. Since
$G_N\mu\sim 10^{-6}$ for GUT strings we can assume that $G_N\mu\ll 1$
and the linear approximation (5.1) is justified. In fact, ignoring quantum
effects the stress tensor (5.5) leads in the linear approximation to the
conical metric with a deficit angle of $8\pi G_N\mu$ \cite{Vil81}

Equations (5.4) can be expressed in terms of the string stress tensor and
its trace $T\equiv T_c= -2\mu\delta (x)\delta (y)$, as,
\begin{eqnarray}
         {B_{(1)\mu}}^\nu (x)
     &=&  16\pi G_N({\delta_\mu}^\nu {T_{,\alpha}}^\alpha
                 -{T_{,\mu}}^\nu),  \nonumber \\
         {A_{(1)\mu}}^\nu (x)
     &=&  8\pi G_N(-2{{{T_\mu}^\nu}_{,\alpha}}^\alpha
          -{2\over 3} {T_{,\mu}}^\nu
          +{2\over 3}{\delta_\mu}^\nu {T_{,\alpha}}^\alpha).
\end{eqnarray}
In this case, ${B_{(1)\mu}}^\nu (x)$ in (5.3) gives  no contribution
outside the core
of the string because it is proportional to partial derivatives of a delta
function with support on the $z$-axis. The tensor components
${A_{(1)\mu}}^\nu (x)$, in the Minkowskian coordinates $(t,x,y,z)$ in
which equation (5.1) is given, are
$$
   {A_{(1)\mu}}^\nu = {8\pi G_N\over 3}\left({\matrix{
             -T_{,xx} -T_{,yy}&0&0&0\cr
             0&\, 2T_{,yy}&-2T_{,xy}&0\cr
             0&-2T_{,yx}&\, 2T_{,xx}&0\cr
             0&0&0&-T_{,xx} -T_{,yy}\cr}}
             \right).
$$
The non-local term $\mbox{\rm H}(x-y;\bar\mu)$ in equation (5.3) can
be expressed as a delta function; following Jordan \cite{Jor87} we can write
\begin{equation}
       \mbox{\rm H}(x-y;\bar\mu) = -{1\over 2\pi }\delta '[(x-y)^2].
\end{equation}

Introducing these expressions into (5.3), we may perform the space integrals;
this is easy due to the presence of the delta functions. In fact, as an
example, let us compute the $x\, y$ component of the stress tensor,
$\langle T^{\;\, y}_x \rangle = \langle T^{\;\, x}_y \rangle$,
to first order in $G_N\mu$.
The only contribution to this component comes from ${A_{(1)x}}^y
= -16\pi G_NT_{,xy}/3=32\pi G_N\mu\delta '(x)\delta '(y)/3$,
\begin{eqnarray}
   \langle T^{\;\, y}_x \rangle &=&3\alpha \int d^4x'
                             \mbox{\rm H}(x-x';\bar\mu)
                             {A_{(1)x}}^y(x')\nonumber \\
                         &=&-{3\alpha\over 2\pi }\int d^4x'\delta '[(x-x')^2]
                             {A_{(1)x}}^y(x').
\end{eqnarray}
This expression can be written as,
\begin{equation}
 \langle T^{\;\, y}_x \rangle
                  = {3\alpha\over 2\pi }\lim_{\lambda\rightarrow 0^-}
                  {d\over d\lambda}\int d^4x'\delta [x'^2-\lambda]
                             {A_{(1)x}}^y(x'^\mu  + x^\mu ),
\end{equation}
where ${A_{(1)x}}^y(x'^\mu  + x^\mu )$ is now
\begin{equation}
      {A_{(1)x}}^y(x'^\mu  + x^\mu )=\left( {32\pi\over 3}\right)G_N\mu
             {\partial\over\partial x}{\partial\over\partial y}
             \left[ \delta (x'+x)\delta (y'+y)\right].
\end{equation}
Following straightforward steps,
\begin{eqnarray}
    \langle T^{\;\, y}_x \rangle &=& 16\alpha G_N\mu
             {\partial\over\partial x}{\partial\over\partial y}
             \left[ \lim_{\lambda\rightarrow 0^-}
                  {d\over d\lambda}\int d^4x'\delta [x'^2-\lambda]
                  \delta (x'+x)\delta (y'+y)\right] \nonumber \\
          &=&16\alpha G_N\mu
             {\partial\over\partial x}{\partial\over\partial y}
             \left[ \lim_{\lambda\rightarrow 0^-}
                  {d\over d\lambda}\int dz'
                  {1\over\sqrt{z'^2+x^2+y^2-\lambda}}\right]
                  \nonumber \\
          &=&16\alpha G_N\mu
             {\partial\over\partial x}{\partial\over\partial y}
             \left[ \int_0^\infty {dz'\over (z'^2+x^2+y^2)^{3/2}}\right]
                  \nonumber \\
          &=&32\alpha G_N\mu\left[ {4xy\over (x^2+y^2)^3}\right] .
\end{eqnarray}
The remaining non null components of the stress tensor in Minkowskian
coordinates to the same order $G_N\mu$ are,
\begin{eqnarray}
   \langle T^{\;\, y}_y \rangle
           &=& 32\alpha G_N\mu\left[
                  {y^2-3x^2\over (x^2+y^2)^3}\right], \nonumber \\
   \langle T^{\;\, x}_x \rangle
           &=& 32\alpha G_N\mu\left[
                  {x^2-3y^2\over (x^2+y^2)^3}\right], \nonumber \\
   \langle T^{\;\, t}_t \rangle
           &=& \langle T^{\;\, z}_z \rangle \, = \, 32\alpha G_N\mu\left[
                  {1\over (x^2+y^2)^2}\right].
\end{eqnarray}
Because of the cylindrical symmetry of the problem
it is better to express this tensor components in polar coordinates as,
\begin{equation}
    \langle {T_\mu}^\nu \rangle \, =
                  \, {32\alpha G_N\mu\over r^4}{\it diag}(1,1,-3,1),
\end{equation}
which is independent of the arbitrary renormalization scale
$\bar\mu $, (as expected because it would contribute with delta functions
with support on the core of the string which we do not consider).
This tensor coincides with the first
order development in $G_N\mu$ of previous
exact results obtained by other techniques \cite{LHK}.
Note that in the exact case, $i.e.$ when the classical solution is found
explicitly, the one loop quantum stress tensor is simply (5.13) where
one changes $32G_N\mu$ by $2[(1-4G_N\mu )^{-4}-1]$.

The backreaction equations (5.2) have been solved by Hiscock \cite{His87}
who found that the linear corrections to the metric outside the
string are such that the spacetime is no longer
flat space with a deficit angle: the two surface perpendicular to the string
is an hyperboloid which asymptotically approaches the conical surface at
large distances (the one loop quantum corrections to $h_{\mu\nu}(x)$
are of the form $G_N\mu\hbar /r^2$). Note that the semiclassical equations
here have no higher order derivatives because we have treated the quantum
terms as a perturbative correction (as Hiscock does) in line with
Simon's arguments \cite{Sim91}. Work on the backreaction
on dynamic cosmic strings \cite{GHV} is in progress.

\section{Acknowledgements}

We would like to thank Bei-Lok Hu and Esteban Calzetta for stimulating
discussions and helpful suggestions. This work has been partially supported
by a CICYT research project number AEN90-0028.

\section{Appendix}

\newcounter{appendix}\def\theappendix{\Alph{appendix}}
\def\theequation{\Alph{appendix}.\arabic{equation}}

\vskip1cm\addtocounter{appendix}{1}
\appendix{\mbox{\LARGE\bf\theappendix}.
          \mbox{\bf Useful relations}}
\setcounter{equation}{0}

\vskip1cm
\appendix{\mbox{\bf\theappendix .1}. Bianchi identities}
\vskip0.5cm

\begin{equation}
      R^{\mu\alpha\nu\beta ;\sigma}
     +R^{\mu\alpha\beta\sigma ;\nu}
     +R^{\mu\alpha\sigma\nu ;\beta}=0.
\end{equation}

\begin{equation}
      {R^{\mu\alpha\nu\beta}}_{;\alpha}=R^{\mu\nu ;\beta}-R^{\mu\beta ;\nu}.
\end{equation}

\begin{equation}
      2{R^{\alpha\beta}}_{;\alpha}=R^{;\beta}.
\end{equation}

\begin{equation}
      {R^{\mu\alpha\nu\beta}}_{;\alpha\beta}=\Box_g R^{\mu\nu}
                  -{R^{\mu\alpha ;\nu}}_\alpha .
\end{equation}

\begin{equation}
      2{R^{\alpha\beta}}_{;\alpha\beta}=\Box_g R.
\end{equation}

\vskip1cm
\appendix{\mbox{\bf\theappendix .2}. Commutation of covariant derivatives}
\vskip0.5cm

\begin{equation}
     {A^{a_1\cdots a_n}}_{;\alpha\beta}
        -{A^{a_1\cdots a_n}}_{;\beta\alpha}=
        -\sum^n_{k=1}{R^{a_k}}_{b_k\alpha\beta}A^{a_1\cdots b_k\cdots a_n}.
\end{equation}

\begin{eqnarray}
     {R^{\mu\alpha ;\nu}}_\alpha
         &=& g^{\nu\alpha}{R^{\mu\beta}}_{;\beta\alpha}
             -R^{\mu\alpha\nu\beta}R_{\alpha\beta}
             +R^{\mu\alpha}{R_\alpha}^\nu
                      \nonumber \\
         &=& {1\over 2}R^{;\mu\nu}
             -R^{\mu\alpha\nu\beta}R_{\alpha\beta}
             +R^{\mu\alpha}{R_\alpha}^\nu .
\end{eqnarray}

\begin{eqnarray}
    {R^{\mu\alpha\nu\beta}}_{;\beta\alpha}
         &=& {R^{\mu\alpha\nu\beta}}_{;\alpha\beta}
             =\Box_g R^{\mu\nu}-{R^{\mu\alpha ;\nu}}_\alpha =
                      \nonumber \\
         &=& \Box_g R^{\mu\nu}-{1\over 2}R^{;\mu\nu}
             +R^{\mu\alpha\nu\beta}R_{\alpha\beta}
             -R^{\mu\alpha}{R_\alpha}^\nu .
\end{eqnarray}

\vskip1cm
\appendix{\mbox{\bf\theappendix .3}. 2D curvature tensors}
\vskip0.5cm

\begin{equation}
      R_{\mu\alpha\nu\beta}=
          {R\over 2}(g_{\mu\nu}g_{\alpha\beta}-g_{\mu\beta}g_{\nu\alpha}).
\end{equation}

\begin{equation}
      R_{\mu\nu}={R\over 2}g_{\mu\nu}.
\end{equation}

\begin{equation}
      G_{\mu\nu}=0.
\end{equation}

\begin{equation}
      R_{\mu\alpha\nu\beta}R^{\mu\alpha\nu\beta}=R^2=2R_{\mu\nu}R^{\mu\nu}.
\end{equation}

\vskip2cm\addtocounter{appendix}{1}
\appendix{\mbox{\LARGE\bf\theappendix}.
          \mbox{\bf Expansions around flat space}}
\setcounter{equation}{0}

\vskip1cm
\appendix{\mbox{\bf\theappendix .1}. Curvature tensors}
\vskip0.5cm

\begin{equation}
           g_{\mu\nu}(x)=\eta_{\mu\nu}+h_{\mu\nu}(x),\,\,\,\,
           (-,+,\cdots ,+).
\end{equation}

\begin{equation}
           g^{\mu\nu}(x)=\eta^{\mu\nu}-h^{\mu\nu}(x)+{h^\mu}_\alpha (x)
                  h^{\alpha\nu}(x)+ O(h^3_{\mu\nu}).
\end{equation}

\begin{equation}
     (-g(x))^{1/2}=
          1+{1\over 2}h+{1\over 8}h^2-{1\over 4}h_{\mu\nu}h^{\mu\nu}
          +O(h^3_{\mu\nu}).
\end{equation}

\begin{equation}
           {R^\alpha}_{\beta\gamma\delta}=
              \partial_\gamma\Gamma^\alpha_{\beta\delta}
             -\partial_\delta\Gamma^\alpha_{\beta\gamma}
             +\Gamma^\alpha_{\mu\gamma}\Gamma^\mu_{\beta\delta}
             -\Gamma^\alpha_{\mu\delta}\Gamma^\mu_{\beta\gamma}
\end{equation}

\begin{equation}
          \Gamma^\alpha_{\beta\delta}=
              {1\over 2}\eta^{\alpha\lambda}S_{\lambda\delta ,\beta}
             -{1\over 2}h^{\alpha\lambda}S_{\lambda\delta ,\beta}
             +O(h^3_{\mu\nu}).
\end{equation}

\begin{equation}
          S_{\lambda\delta ,\beta}\equiv
              h_{\lambda\delta ,\beta}
             +h_{\beta\lambda ,\delta}
             -h_{\beta\delta ,\lambda}.
\end{equation}

\begin{equation}
          R_{\sigma\beta\gamma\delta}=
                   S_{\sigma [\delta ,\beta\gamma ]}
                   -{h^\lambda}_{\sigma ,[\gamma}S_{\lambda\delta ],\beta}
                   +{1\over 2}\eta^{\mu\rho}
                   S_{\sigma [\gamma ,\mu}S_{\rho\delta ],\beta}
                   +O(h^3_{\mu\nu}).
\end{equation}

\begin{eqnarray}
          R_{\beta\delta}
             &=&   {1\over 2}\eta^{\mu\nu}
                   (h_{\mu\delta ,\beta\nu}
                   -h_{\beta\delta ,\mu\nu}
                   -h_{\mu\nu ,\beta\delta}
                   +h_{\beta\nu ,\mu\delta})
                   \nonumber \\
             & &   -\left( {h^{\mu\nu}}_{,\mu}-{1\over 2}h^{,\nu}\right)
                   h_{\nu (\delta ,\beta )}
                   +{1\over 2}(h^{\mu\nu}h_{\beta\delta ,\nu})_{,\mu}
                   +{1\over 4}{h^{\mu\nu}}_{,\delta}h_{\mu\nu ,\beta}
                   -{1\over 4}h^{,\mu}h_{\beta\delta ,\mu}
                   \nonumber \\
             & &   +h_{\mu\delta ,\nu}{h_\beta}^{[\mu ,\nu ]}
                   +{1\over 2}h^{\mu\nu}h_{\mu\nu ,\beta\delta}
                   -h^{\mu\nu}h_{\mu (\beta ,\delta )\nu}
                   +O(h^3_{\mu\nu})
\end{eqnarray}

\begin{eqnarray}
          R  &=&   {h^{\alpha\beta}}_{,\alpha\beta}
                   -{h_{,\alpha}}^{\alpha}
                   \nonumber \\
             & &  - {h^{\mu\nu}}_{,\mu} {h_{\nu\alpha}}^{,\alpha}
                  + {h^{\mu\nu}}_{,\mu} h_{,\nu}
                  + {3\over 4}h^{\mu\nu ,\alpha}h_{\mu\nu ,\alpha}
                  - {1\over 4}h^{,\mu}h_{,\mu}
                   \nonumber \\
             & &  - {1\over 2}h^{\mu\nu ,\alpha}h_{\alpha\mu ,\nu}
                  -2h^{\mu\nu}{h_{\mu\alpha ,\nu}}^\alpha
                  +h^{\mu\nu}h_{,\mu\nu}
                  +h^{\mu\nu}{h_{\mu\nu ,\alpha}}^\alpha
                   +O(h^3_{\mu\nu}).
\end{eqnarray}

\begin{equation}
            R_{\sigma\beta\gamma\delta} R^{\sigma\beta\gamma\delta}=
                h_{\sigma\delta ,\beta\gamma}h^{\sigma\delta ,\beta\gamma}
              -2h_{\sigma\delta ,\beta\gamma}h^{\beta\delta ,\sigma\gamma}
               +h_{\sigma\delta ,\beta\gamma}h^{\beta\gamma ,\sigma\delta}
               +O(h^3_{\mu\nu}).
\end{equation}

\begin{eqnarray}
  R_{\beta\delta} R^{\beta\delta} &=&
    {1\over 4}\left[\atop \right.
    2{h^{\mu\nu}}_{,\alpha\mu}{h_{\beta\nu}}^{,\alpha\beta}
   -4{h^{\mu\nu}}_{,\alpha\mu}{{h^\alpha}_{\nu ,\beta}}^\beta
   -4{h^{\mu\nu}}_{,\alpha\mu}{h^{,\alpha}}_{\nu}
   +2{h^{\mu\nu}}_{,\alpha\mu}{h^{\alpha\beta}}_{,\beta\nu}
     \nonumber \\ & &\hskip1cm
   +{{h^{\mu\nu}}_{,\alpha}}^\alpha {h_{\mu\nu ,\beta}}^\beta
   +2{{h^{\mu\nu}}_{,\alpha}}^\alpha h_{,\mu\nu}
   +h^{,\mu\nu}h_{,\mu\nu}
   +O(h^3_{\mu\nu})\left.\atop\right].
\end{eqnarray}

\begin{equation}
     R^2 =
          {h^{\alpha\beta}}_{,\alpha\beta}{h^{\mu\nu}}_{,\mu\nu}
          -2{h^{\alpha\beta}}_{,\alpha\beta}{h_{,\mu}}^{\mu}
          +{h_{,\alpha}}^{\alpha}{h_{,\mu}}^{\mu}
          +O(h^3_{\mu\nu}).
\end{equation}

\vskip1cm
\appendix{\mbox{\bf\theappendix .2}. Curvature tensors in terms of
                     the projector $P_{\mu\nu}$}
\vskip0.5cm

\begin{equation}
       P_{\mu\nu}=\eta_{\mu\nu}-{q^\mu q^\nu\over q^2}.
\end{equation}
\begin{equation}
     G\equiv\int d^nx\left(
         R_{\mu\nu\alpha\beta} R^{\mu\nu\alpha\beta}
       -4R_{\mu\nu} R^{\mu\nu}+R^2\right)=0 +O(h^3_{\mu\nu}).
\end{equation}

\begin{equation}
     \int d^nxR^2(x)=\int d^nx d^ny h_{\mu\nu}(x)h_{\alpha\beta}(y)
                \int {d^nq\over(2\pi )^n}e^{iq\cdot (x-y)}
                P^{\mu\nu}P^{\alpha\beta}q^4
                +O(h^3_{\mu\nu}).
\end{equation}

\begin{equation}
     \int d^nxR_{\mu\nu\alpha\beta}(x)R^{\mu\nu\alpha\beta}(x)
                =\int d^nx d^ny h_{\mu\nu}(x)h_{\alpha\beta}(y)
                \int {d^nq\over(2\pi )^n}e^{iq\cdot (x-y)}
                P^{\mu\beta}P^{\nu\alpha}q^4
                +O(h^3_{\mu\nu}).
\end{equation}

\begin{eqnarray}
     \int d^nxd^nyR(x)R(y)\mbox{\rm K}_1(x-y;\mu)
                &=&-{1\over2}\int d^nx d^ny h_{\mu\nu}(x)h_{\alpha\beta}(y)
                \nonumber \\ & &\hskip-2.3cm \times
                \int {d^nq\over(2\pi )^n}e^{iq\cdot (x-y)}
                \ln\left[{q^2-i\epsilon\over\mu^2}\right]
                P^{\mu\nu}P^{\alpha\beta}q^4
                +O(h^3_{\mu\nu}).
\end{eqnarray}

\begin{eqnarray}
     \int d^nx d^nyR_{\mu\nu\alpha\beta}(x)R^{\mu\nu\alpha\beta}(y)
                \mbox{\rm K}_1(x-y;\mu)
                &=&-{1\over2}\int d^nx d^ny h_{\mu\nu}(x)h_{\alpha\beta}(y)
                \nonumber \\  & &\hskip-4cm \times
                \int {d^nq\over(2\pi )^n}e^{iq\cdot (x-y)}
                \ln\left[{q^2-i\epsilon\over\mu^2}\right]
                P^{\mu\beta}P^{\nu\alpha}q^4
                +O(h^3_{\mu\nu}).
\end{eqnarray}

\begin{eqnarray}
     \int d^nx d^nyR(x)R(y)\mbox{\rm K}_2(x-y)
                &=&-{1\over2}\int d^nx d^ny h_{\mu\nu}(x)h_{\alpha\beta}(y)
                \nonumber \\  & &\hskip-3cm \times
                \int {d^nq\over(2\pi )^n}e^{iq\cdot (x-y)}
                (2\pi i)\theta (-q^2)\theta (-q^0)
                P^{\mu\nu}P^{\alpha\beta}q^4
                +O(h^3_{\mu\nu}).
\end{eqnarray}

\begin{eqnarray}
     \int d^nx d^nyR_{\mu\nu\alpha\beta}(x)R^{\mu\nu\alpha\beta}(y)
                \mbox{\rm K}_2(x-y)
                &=&-{1\over2}\int d^nx d^ny h_{\mu\nu}(x)h_{\alpha\beta}(y)
                \nonumber \\  & &\hskip-5cm \times
                \int {d^nq\over(2\pi )^n}e^{iq\cdot (x-y)}
                (2\pi i)\theta (-q^2)\theta (-q^0)
                P^{\mu\beta}P^{\nu\alpha}q^4
                +O(h^3_{\mu\nu}).
\end{eqnarray}

\vskip2cm\addtocounter{appendix}{1}
\appendix{\mbox{\LARGE\bf\theappendix}.
          \mbox{\bf Curvature tensors in conformally transformed
                    n diamensional spaces}}
\setcounter{equation}{0}
\vskip1cm

\begin{equation}
    \tilde g_{\mu\nu}(x)=e^{2\omega}g_{\mu\nu}(x)
\end{equation}

\begin{eqnarray}
    \tilde R_{\mu\nu\alpha\beta}=
         e^{2\omega}\left[\right.
         & &\hskip -0.8cm
             R_{\mu\nu\alpha\beta}
            +2g_{\beta [\mu}{\delta_{\nu ]}}^\rho
                (\omega_{;\alpha\rho}-\omega_{;\alpha}\omega_{;\rho})
            -2g_{\alpha [\mu}{\delta_{\nu ]}}^\rho
                (\omega_{;\beta\rho}-\omega_{;\beta}\omega_{;\rho})
            \nonumber \\
         & &\hskip -0.8cm
            -2g_{\alpha [\mu}g_{\beta\nu ]}\omega_{;\rho}\omega^{;\rho}
                \left.\right].
\end{eqnarray}

\begin{equation}
   \tilde R_{\mu\alpha}=R_{\mu\alpha}-(n-2)\omega_{;\mu\alpha}
            -g_{\mu\alpha}\Box_g\omega +(n-2)\omega_{;\mu}\omega_{;\alpha}
            -(n-2)g_{\mu\alpha}\omega_{;\rho}\omega^{;\rho}.
\end{equation}

\begin{equation}
   \tilde R=e^{-2\omega}\left[ R-2(n-1)\Box_g\omega -(n-1)(n-2)
               \omega_{;\rho}\omega^{;\rho}\right].
\end{equation}

\begin{eqnarray}
    \tilde R_{\mu\nu\alpha\beta} \tilde R^{\mu\nu\alpha\beta}=e^{-4\omega}
            \left[\right.
        & & \hskip -0.8cm
             R_{\mu\nu\alpha\beta}R^{\mu\nu\alpha\beta}
            -8R^{\mu\nu}(\omega_{;\mu\nu}-\omega_{;\mu}\omega_{;\nu})
            -4R\omega_{;\mu}\omega^{;\mu}
            \nonumber \\
        & & \hskip -0.8cm
            +4(n-2)\omega_{;\mu\nu}\omega^{;\mu\nu}
            -8(n-2)\omega^{;\mu\nu}\omega_{;\mu}\omega_{;\nu}
            +4(\Box_g\omega )^2
            \nonumber \\
        & & \hskip -0.8cm
            +8(n-2)\omega_{;\mu}\omega^{;\mu}\Box_g\omega
            +2(n-1)(n-2)(\omega_{;\mu}\omega^{;\mu})^2
            \left.\right].
\end{eqnarray}

\begin{eqnarray}
    \tilde R_{\mu\nu} \tilde R^{\mu\nu}=e^{-4\omega}
            \left[\right.
         & &\hskip -0.8cm
            R_{\mu\nu}R^{\mu\nu}
            -2(n-2)R^{\mu\nu}(\omega_{;\mu\nu}-\omega_{;\mu}\omega_{;\nu})
            -2R\Box_g\omega
            \nonumber \\
         & &\hskip -0.8cm
            -2(n-2)R\omega_{;\mu}\omega^{;\mu}
            +(n-2)^2\omega_{;\mu\nu}\omega^{;\mu\nu}
            +(3n-4)(\Box_g\omega )^2
            \nonumber \\
         & &\hskip -0.8cm
            -2(n-2)^2\omega^{;\mu\nu}\omega_{;\mu}\omega_{;\nu}
            +2(n-2)(2n-3)\omega_{;\mu}\omega^{;\mu}\Box_g\omega
            \nonumber \\
         & &\hskip -0.8cm
            +(n-1)(n-2)^2(\omega_{;\mu}\omega^{;\mu})^2
            \left.\right].
\end{eqnarray}

\begin{eqnarray}
    \tilde R^2=e^{-4\omega}
            \left[\right.
         & &\hskip -0.8cm
             R^2
            -4(n-1)R\Box_g\omega
            -2(n-1)(n-2)R\omega_{;\mu}\omega^{;\mu}
            +4(n-1)^2(\Box_g\omega )^2
            \nonumber \\
         & &\hskip -0.8cm
            +4(n-1)^2(n-2)\omega_{;\mu}\omega^{;\mu}\Box_g\omega
            +(n-1)^2(n-2)^2(\omega_{;\mu}\omega^{;\mu})^2
            \left.\right].
\end{eqnarray}

\begin{eqnarray}
     \tilde C_{\mu\nu\alpha\beta} \tilde C^{\mu\nu\alpha\beta}&=&
             \tilde R_{\mu\nu\alpha\beta} \tilde R^{\mu\nu\alpha\beta}
             -{4\over (n-2)} \tilde R_{\mu\nu}\tilde R^{\mu\nu}
             +{2\over (n-1)(n-2)}\tilde R^2
             \nonumber \\
         &=& e^{-4\omega}C_{\mu\nu\alpha\beta}C^{\mu\nu\alpha\beta}
\end{eqnarray}

\vskip2cm\addtocounter{appendix}{1}
\appendix{\mbox{\LARGE\bf\theappendix}.
          \mbox{\bf Momentum integrals and dimensional regularization}}
\setcounter{equation}{0}
\vskip1cm

\begin{equation}
     I(q)\equiv\int {d^np\over (2\pi )^n}f(p,q).
\end{equation}

\begin{equation}
     I_\mu =\int {d^np\over (2\pi )^n}f(p,q)p_\mu = {I(q)\over 2}q_\mu.
\end{equation}

\begin{equation}
     I_{\mu\nu}=\int {d^np\over (2\pi )^n}f(p,q)p_\mu p_\nu
               ={I(q)\over 4}\left[q_\mu q_\nu
                  - {q^2\over (n-1)}P_{\mu\nu}\right].
\end{equation}

\begin{eqnarray}
     I_{\mu\nu\alpha}&=&\int {d^np\over (2\pi )^n}f(p,q)p_\mu p_\nu p_\alpha
                        \nonumber \\
                     &=&{I(q)\over 8}\left[q_\mu q_\nu q_\alpha
                         - {q^2\over (n-1)}
                        (P_{\mu\nu}q_\alpha +P_{\mu\alpha}q_\nu
                        +P_{\alpha\nu}q_\mu )\right].
\end{eqnarray}

\begin{eqnarray}
     I_{\mu\nu\alpha\beta}
         &=&\int {d^np\over (2\pi )^n}f(p,q)p_\mu p_\nu p_\alpha p_\beta
                        \nonumber \\
         &=&{I(q)\over 16}\left\{\atop\right. q_\mu q_\nu q_\alpha q_\beta
                  - {q^2\over (n-1)}
                        \left[ P_{\mu\nu}q_\alpha q_\beta
                        +P_{\nu\alpha}q_\mu q_\beta
                        +P_{\nu\beta}q_\mu q_\alpha\right.
                        \nonumber \\
             & &\hskip 4.5cm
                        \left. +P_{\mu\alpha}q_\nu q_\beta
                        +P_{\mu\beta}q_\nu q_\alpha
                        +P_{\alpha\beta}q_\mu q_\nu\right]
                        \nonumber \\
             & &\hskip 3cm
                  + {q^4\over (n^2-1)}
                        \left[ P_{\mu\nu}P_{\alpha\beta}
                        +P_{\mu\beta}P_{\nu\alpha}
                        +P_{\nu\beta}P_{\mu\alpha}
                        \right]
                     \left.\atop\right\}.
\end{eqnarray}

\begin{eqnarray}
       I_1(p)&=&\int {d^nq\over (2\pi )^n}
               {1\over (q^2-i\epsilon )[(p-q)^2-i\epsilon ]}
               \nonumber \\
             &=&i(p^2-i\epsilon )^{{n\over 2}-2}
               {\Gamma \left( 2-{n\over 2}\right)
                  \left[\Gamma \left( {n\over 2}-1\right)\right]^2
                 \over (4\pi )^{n\over 2}\Gamma (n-2)}.
\end{eqnarray}

\begin{equation}
       I^{(4D)}_1(p)=\left( {-i\over 8\pi^2}\right)
            \left[ {1\over n-4}+{1\over 2}\ln (p^2-i\epsilon )
                   +{1\over 2}\left( \gamma -2 -\ln 4\pi \right)+
                   O(n-4)\right].
\end{equation}

\begin{equation}
       I^{(2D)}_1(p)={i\over \pi (p^2-i\epsilon )}
            \left[ {1\over n-2}+{1\over 2} \gamma +O(n-2)\right].
\end{equation}

\begin{eqnarray}
       I_2(p)&=&\int {d^nq\over (2\pi )^n}
               \delta (q^2)\theta (-q^0)\delta [(p-q)^2]\theta (q^0-p^0)
                \nonumber \\
             &=&{\theta (-p^0)\theta (-p^2)(p^2)^{n-3}
                \over (n-3)2^{n-1}(2\pi )^3|\vec p|}
                \left[ {1\over (p^0+|\vec p|)^{n-3}}
                      -{1\over (p^0-|\vec p|)^{n-3}}\right].
\end{eqnarray}

\begin{equation}
       I^{(4D)}_2(p)={\theta (-p^0)\theta (-p^2)\over 4(2\pi )^3}.
\end{equation}

\begin{equation}
       I^{(2D)}_2(p)=-{\theta (-p^0)\theta (-p^2)
                        \over (2\pi )^3(p^2-i\epsilon )}.
\end{equation}

\vskip2cm\addtocounter{appendix}{1}
\appendix{\mbox{\LARGE\bf\theappendix}.
          \mbox{\bf Variational calculus}}
\setcounter{equation}{0}

\vskip1cm
\appendix{\mbox{\bf\theappendix .1}. Variational equations}
\vskip0.5cm

\begin{equation}
     g_{\mu\nu}(x)\longrightarrow g_{\mu\nu}(x)+\delta g_{\mu\nu}(x).
\end{equation}

\begin{equation}
     \delta g^{\alpha\lambda} = -g^{\alpha\mu}g^{\lambda\nu}\delta g_{\mu\nu}.
\end{equation}

\begin{equation}
     \delta \sqrt{-g}={1\over 2}\sqrt{-g}g^{\mu\nu}\delta g_{\mu\nu}.
\end{equation}

\begin{equation}
     \delta\Gamma^\alpha_{\beta\delta}={1\over 2}g^{\alpha\lambda}
            \left[ \delta g_{\lambda\delta ;\beta}
                  +\delta g_{\beta\lambda ;\delta}
                  -\delta g_{\beta\delta ;\lambda}\right].
\end{equation}

\begin{equation}
     \delta {R^\alpha}_{\beta\gamma\delta}={1\over 2}g^{\alpha\lambda}
            \left[ \delta g_{\lambda\delta ;\beta\gamma}
                  +\delta g_{\beta\lambda ;\delta\gamma}
                  -\delta g_{\beta\delta ;\lambda\gamma}
                  -\delta g_{\lambda\gamma ;\beta\delta}
                  -\delta g_{\beta\lambda ;\gamma\delta}
                  +\delta g_{\beta\gamma ;\lambda\delta}\right].
\end{equation}

\begin{equation}
     \delta R_{\beta\delta}={1\over 2}g^{\alpha\lambda}
            \left[ \delta g_{\lambda\delta ;\beta\alpha}
                  +\delta g_{\beta\lambda ;\delta\alpha}
                  -\delta g_{\beta\delta ;\lambda\alpha}
                  -\delta g_{\lambda\alpha ;\beta\delta}\right].
\end{equation}

\begin{equation}
     \delta R=-R^{\beta\delta}\delta g_{\beta\delta}
            +g^{\beta\delta}g^{\alpha\lambda}
            \left[ \delta g_{\lambda\delta ;\beta\alpha}
                  -\delta g_{\beta\delta ;\lambda\alpha}\right].
\end{equation}

\begin{equation}
     \delta (\Box_g\omega)=-\omega^{;\mu\nu}\delta g_{\mu\nu}
            -{1\over 2}\omega^{;\lambda}g^{\mu\nu}
            \left[ \delta g_{\lambda\nu ;\mu}
                  +\delta g_{\mu\lambda ;\nu}
                  -\delta g_{\mu\nu ;\lambda}\right].
\end{equation}

\vskip1cm
\appendix{\mbox{\bf\theappendix .2}. Functional differentiation}
\vskip0.5cm

\begin{equation}
     \delta\int d^4x\sqrt{-g}R^2=\int d^4x\sqrt{-g}
           \left\{ {1\over 2}g^{\mu\nu}R^2
                 -2RR^{\mu\nu}+2R^{;\mu\nu}-2g^{\mu\nu}\Box_g R
           \right\}\delta g_{\mu\nu}.
\end{equation}

\begin{eqnarray}
     \hskip -1cm\delta\int d^4x\sqrt{-g}R^{\alpha\beta}R_{\alpha\beta}
           \hskip-0.3cm&=&\hskip-0.3cm
           \int d^4x\sqrt{-g}
           \left\{ {1\over 2}g^{\mu\nu}R^{\alpha\beta}R_{\alpha\beta}
                 -2R^{\mu\alpha}{R_\alpha}^\nu
                 +2{R^{\mu\alpha ;\nu}}_\alpha
                 -\Box_g R^{\mu\nu}
                 \right.\nonumber \\ & & \hskip 1.7cm\left.
                 -{1\over 2}g^{\mu\nu}\Box_g R
           \right\}\delta g_{\mu\nu}.
\end{eqnarray}

\begin{eqnarray}
     \hskip-1cm\delta\int d^4x\sqrt{-g}R^{\alpha\beta\rho\sigma}
           R_{\alpha\beta\rho\sigma}
           \hskip-0.3cm&=&\hskip-0.3cm\int d^4x\sqrt{-g}
           \left\{ {1\over 2}g^{\mu\nu}R^{\alpha\beta\rho\sigma}
                  R_{\alpha\beta\rho\sigma}
                 -2R^{\mu\alpha\beta\rho}{R^\nu}_{\alpha\beta\rho}
                 -4\Box_g R^{\mu\nu}
                 \right.\nonumber \\ & &\hskip 2cm
                 +2R^{;\mu\nu}
                 -4R^{\mu\alpha\nu\beta}R_{\alpha\beta}
                 +4R^{\mu\alpha}{R_\alpha}^\nu \left.{\atop}
           \right\}\delta g_{\mu\nu}.
\end{eqnarray}

\begin{eqnarray}
     \hskip-2em\delta\int d^4x\sqrt{-g}
           R^{\alpha\beta}\omega_{;\alpha}\omega_{;\beta}
           \hskip-0.3cm&=&\hskip-0.3cm\int d^4x\sqrt{-g}
           \left\{ {\atop}g^{\mu\nu}
           R^{\alpha\beta}\omega_{;\alpha}\omega_{;\beta}
           -2R^{(\mu\alpha}\omega^{;\nu )}\omega_{;\alpha}
           +\left(\Box_g\omega\right)\omega^{;\mu\nu}
           \right.\nonumber \\ & &\hskip-1cm\left.
           -{1\over 2}g^{\mu\nu}\left[ \left(\Box_g\omega\right)^2
           +\omega^{;\alpha\beta}\omega_{;\alpha\beta}\right]
           +\left[\omega^{;\mu\nu\alpha}
           -g^{\mu\nu}\Box_g\left(\omega^{;\alpha}\right)\right]
           \omega_{;\alpha}{\atop}\right\}\delta g_{\mu\nu}.
\end{eqnarray}

\begin{eqnarray}
     \hskip -1cm\delta\int d^4x\sqrt{-g}R\left(\Box_g\omega\right)
           \hskip-0.3cm&=&\hskip-0.3cm\int d^4x\sqrt{-g}\left\{
           -R^{\mu\nu}\left(\Box_g\omega\right)
           +R^{;(\mu}\omega^{;\nu )}
           -{1\over 2}g^{\mu\nu}R^{;\alpha}\omega_{;\alpha}
           +\left(\Box_g\omega\right)^{;\mu\nu}
           \right.\nonumber \\ & &\hskip 2cm \left.
           -g^{\mu\nu}\Box_g\left(\Box_g\omega\right)
           {\atop} \right\}\delta g_{\mu\nu}.
\end{eqnarray}

\begin{eqnarray}
     \hskip-0.5cm\delta\int d^4x\sqrt{-g}
           R\left(\omega_{;\alpha}\omega^{;\alpha}\right)
           \hskip-0.3cm&=&\hskip-0.3cm\int d^4x\sqrt{-g}\left\{
           {1\over 2}g^{\mu\nu}R\left(\omega_{;\alpha}\omega^{;\alpha}\right)
           -R^{\mu\nu}\left(\omega_{;\alpha}\omega^{;\alpha}\right)
           -R\left(\omega^{;\mu}\omega^{;\nu}\right)
           \right.\nonumber \\ & &\hskip 2cm
           +2R^{\mu\alpha\nu\beta}\omega_{;\alpha}\omega_{;\beta}
           +2\left[\omega^{;\mu\alpha}{\omega^{;\nu}}_\alpha
           -g^{\mu\nu}\omega_{;\alpha\beta}\omega^{;\alpha\beta}\right]
           \nonumber \\ & &\hskip 2cm\left.
           +2\left[\omega^{;\mu\nu\alpha}
           -g^{\mu\nu}\Box_g(\omega^{;\alpha})\right]\omega_{;\alpha}
           {\atop}\right\}\delta g_{\mu\nu}.
\end{eqnarray}

\begin{equation}
     \delta\int d^4x\sqrt{-g}
           \Box_g\omega\left(\omega_{;\alpha}\omega^{;\alpha}\right)
           \hskip-0.1cm =\hskip-0.2cm
           \int d^4x\sqrt{-g}\left\{{\atop}\hskip-0.5em
           \left[2\omega^{;\alpha (\mu}\omega^{;\nu )}
           -g^{\mu\nu}\omega^{;\alpha\beta}\omega_{;\beta}
           \right]\omega_{;\alpha}
           -\left(\Box_g\omega\right)\omega^{;\mu}\omega^{;\nu}
           {\atop}\right\}\delta g_{\mu\nu}.
\end{equation}

\begin{equation}
     \delta\int d^4x\sqrt{-g}\left(\Box_g\omega\right)^2
           =\int d^4x\sqrt{-g}\left\{
           -g^{\mu\nu}\left[{1\over2}\left(\Box_g\omega\right)^2
           +\left(\Box_g\omega\right)_{;\alpha}\omega^{;\alpha}\right]
           +2\omega^{;(\mu}\left(\Box_g\omega\right)^{;\nu )}
           \right\}\delta g_{\mu\nu}.
\end{equation}

\begin{equation}
     \delta\int d^4x\sqrt{-g}\left(\omega_{;\alpha}\omega^{;\alpha}\right)^2
           =\int d^4x\sqrt{-g}\left\{
           {1\over2}g^{\mu\nu}\left(\omega_{;\alpha}\omega^{;\alpha}\right)^2
           -2\omega^{;\mu}\omega^{;\nu}
           \left(\omega_{;\alpha}\omega^{;\alpha}\right)
           \right\}\delta g_{\mu\nu}.
\end{equation}

\begin{eqnarray}
     \hskip-1cm\delta\int d^4x\sqrt{-g}R^{\mu\nu}R_{\mu\nu}\omega(x)
           \hskip-0.3cm&=&\hskip-0.3cm\int d^4x\sqrt{-g}\left\{
           \left[{1\over2}g^{\mu\nu}R^{\alpha\beta}R_{\alpha\beta}
           -2R^{\mu\alpha\nu\beta}R_{\alpha\beta}
           -R^{;\mu\nu}\right.\right.
           \nonumber \\ & &\hskip 1.2cm\left.
           -\Box_g R^{\mu\nu}
           -{1\over2}g^{\mu\nu}\Box_g R\right]\omega(x)
           +R^{;(\mu}\omega^{;\nu)}
           \nonumber \\ & &\hskip 1.2cm
           +\left[2R^{\alpha (\mu;\nu )}
           -2R^{\mu\nu ;\alpha}
           -g^{\mu\nu}R^{;\alpha}\right]\omega_{;\alpha}
           \nonumber \\ & &\hskip 1.2cm\left.
           +2R^{\alpha (\mu}{\omega^{;\nu )}}_\alpha
           -R^{\mu\nu}\left(\Box_g\omega\right)
           -g^{\mu\nu}R^{\alpha\beta}\omega_{;\alpha\beta}
           {\atop}\right\}\delta g_{\mu\nu}.
\end{eqnarray}

\begin{eqnarray}
     \hskip-1cm\delta\int d^4x\sqrt{-g}
           R^{\alpha\beta\rho\sigma}R_{\alpha\beta\rho\sigma}\omega(x)
           \hskip-0.3cm&=&\hskip-0.3cm\int d^4x\sqrt{-g}\left\{
           \left[{1\over2}g^{\mu\nu}
           R^{\alpha\beta\rho\sigma}R_{\alpha\beta\rho\sigma}
           -2R^{\mu\beta\rho\sigma}{R^\nu}_{\beta\rho\sigma}\right.\right.
           \nonumber \\ & &\hskip1cm\left.
           -4R^{\mu\alpha\nu\beta}R_{\alpha\beta}
           +4R^{\mu\alpha}{R_\alpha}^\nu
           -4\Box_g R^{\mu\nu}
           +2R^{;\mu\nu}\right]\omega(x)
           \nonumber \\ & &\hskip1cm\left.
           -8\left[R^{\mu\nu ;\alpha}
           -R^{\alpha (\mu ;\nu )}
           \right]\omega_{;\alpha}
           -4R^{\mu\alpha\nu\beta}\omega_{;\alpha\beta}
           {\atop}\right\}\delta g_{\mu\nu}.
\end{eqnarray}

\begin{eqnarray}
     \hskip -2cm\delta\int d^4xd^4y\sqrt{-g^+(x)}\sqrt{-g^+(y)}
           R^+(x)R^+(y)\mbox{\rm K}_1(x-y;\bar\mu)
           \hskip-0.3cm&=&\nonumber \\ & &\hskip -8cm
           =\int d^4x\sqrt{-g^+(x)}\left\{
           2\int d^4y\sqrt{-g^+(y)}R^+(y)\right.
           \nonumber \\& &\hskip -7cm\left.
           \times\left[
           -G^{+\mu\nu}(x)
           +\nabla^\mu_{(x)}\nabla^\nu_{(x)}
           -g^{+\mu\nu}(x)\Box_{(x)}\right]
           \mbox{\rm K}_1(x-y;\bar\mu)
           {\atop}\right\}\delta g^+_{\mu\nu}(x).
\end{eqnarray}

\begin{eqnarray}
     \hskip -2cm\delta\int d^4xd^4y\sqrt{-g^+(x)}\sqrt{-g^+(y)}
           R^{+\alpha\beta\rho\sigma}(x)R^+_{\alpha\beta\rho\sigma}(y)
           \mbox{\rm K}_1(x-y;\bar\mu)
           \hskip-0.3cm&=&\nonumber \\ & &\hskip -9cm
           =\int d^4x\sqrt{-g^+(x)}\left\{
           \int d^4y\sqrt{-g^+(y)}\left[
           g^{+\mu\nu}(x)R^+_{\alpha\beta\rho\sigma}(x)
           R^{+\alpha\beta\rho\sigma}(y)\right.\right.
           \nonumber \\& &\hskip -5.3cm\left.\left.
           -4R^{+\mu\alpha\nu\beta}(y)
           \nabla^{(x)}_\alpha\nabla^{(x)}_\beta\right]
           \mbox{\rm K}_1(x-y;\bar\mu){\atop}\right\}\delta g^+_{\mu\nu}(x).
\end{eqnarray}

\begin{eqnarray}
     \hskip -1.5cm\delta\int d^4xd^4y\sqrt{-g^+(x)}\sqrt{-g^-(y)}
           R^+(x)R^-(y)\mbox{\rm K}_2(x-y)
           \hskip-0.3cm&=&\nonumber \\ & &\hskip -7cm
           =\int d^4x\sqrt{-g^+(x)}\left\{
           \int d^4y\sqrt{-g^-(y)}R^-(y)\right.
           \nonumber \\& &\hskip -6cm\left.
           \times\left[
           -G^{+\mu\nu}(x)
           +\nabla^\mu_{(x)}\nabla^\nu_{(x)}
           -g^{+\mu\nu}(x)\Box_{(x)}\right]
           \mbox{\rm K}_2(x-y)
           {\atop}\right\}\delta g^+_{\mu\nu}(x).
\end{eqnarray}

\begin{eqnarray}
     \hskip-1cm\delta\int d^4xd^4y\sqrt{-g^+(x)}\sqrt{-g^-(y)}
           R^+_{\alpha\beta\rho\sigma}(x)R^{-\alpha\beta\rho\sigma}(y)
           \mbox{\rm K}_2(x-y)
           \hskip-0.3cm&=&\nonumber \\ & &\hskip -9cm
           =\int d^4x\sqrt{-g^+(x)}\left\{
           \int d^4y\sqrt{-g^-(y)}]\left[{1\over2}
           g^{+\mu\nu}(x)
           R^+_{\alpha\beta\rho\sigma}(x)R^{-\alpha\beta\rho\sigma}(y)
           \right.\right.
           \nonumber \\& &\hskip -4.7cm\left.\left.
           -2R^{-\mu\alpha\nu\beta}(y)
           \nabla^{(x)}_\alpha\nabla^{(x)}_\beta{\atop}\right]
           \mbox{\rm K}_2(x-y)
           {\atop}\right\}\delta g^+_{\mu\nu}(x).
\end{eqnarray}

\end{document}